\documentclass[titlepage,12pt]{article}
\usepackage{amssymb,amsmath,color,graphicx,amscd,epsf,indentfirst,amsfonts}
\usepackage{mathtools}
\usepackage{enumerate}
\usepackage{epsfig}
\usepackage[colorlinks=true,citecolor=blue]{hyperref}
\usepackage{cite}
\usepackage{scalefnt}
\usepackage[titles]{tocloft}
\usepackage{sectsty}
\allsectionsfont{\bf \scalefont{.7} \selectfont}
\subsectionfont{\bf \scalefont{.85} \it \selectfont}
\subsubsectionfont{\bf \scalefont{1} \it \selectfont}
\usepackage[T1]{fontenc}
\usepackage{lmodern}
\usepackage{bibspacing}

\def\blfootnote{\xdef\@thefnmark{}\@footnotetext}

\long\def\symbolfootnote[#1]#2{\begingroup%
\def\thefootnote{\fnsymbol{footnote}}\footnote[#1]{#2}\endgroup}

\makeatletter
\renewcommand{\@dotsep}{4.5}
\makeatother

\def\be{\begin{equation}}
\def\ee{\end{equation}}

\makeatletter
\def\@seccntformat#1{\csname the#1\endcsname.\quad}
\makeatother

\def\clock{{\count0=\time
           \divide\count0 60
           \ifnum\count0<10 0\fi\the\count0
           \multiply\count0 -60 \advance\count0 \time
           :\ifnum\count0<10 0\fi \the\count0
         }}
\newcommand{\timestamp}{{\small\vbox{\hbox{\tt\jobname.tex}
\hbox{\the\day/\the\month/\the\year, \clock}}}}

\def\EE{{\cal E}}
\def\FF{{\cal F}}

\def\NN{{\cal N}}
\def\OO{{\cal O}}

\def\SS{{\cal S}}

\def\ZZ{{\cal Z}}

\def\IR{{\mathbb R}}

\def\tts{{$tt^*$ }}
\def\A{{\cal A}}

\def\Xint#1{\mathchoice
   {\XXint\displaystyle\textstyle{#1}}%
   {\XXint\textstyle\scriptstyle{#1}}%
   {\XXint\scriptstyle\scriptscriptstyle{#1}}%
   {\XXint\scriptscriptstyle\scriptscriptstyle{#1}}%
   \!\int}
\def\XXint#1#2#3{{\setbox0=\hbox{$#1{#2#3}{\int}$}
     \vcenter{\hbox{$#2#3$}}\kern-.5\wd0}}

\def\dashint{\Xint-}

\def\barM{\overline}

\def\Tr{{\rm {Tr}}}

\def\d{{\partial}}
\def\bd{\boldsymbol{\partial}}

\def\beq{\begin{equation}}
\def\eeq{\end{equation}}
\newcommand{\bea}{\begin{eqnarray}}
\newcommand{\eea}{\end{eqnarray}}
\def\bal{\begin{align}}
\def\eal{\end{align}}

\makeatletter
\newsavebox{\@brx}
\newcommand{\llangle}[1][]{\savebox{\@brx}{\(\m@th{#1\langle}\)}%
  \mathopen{\copy\@brx\kern-0.5\wd\@brx\usebox{\@brx}}}
\newcommand{\rrangle}[1][]{\savebox{\@brx}{\(\m@th{#1\rangle}\)}%
  \mathclose{\copy\@brx\kern-0.5\wd\@brx\usebox{\@brx}}}
\makeatother

\numberwithin{equation}{section}

\begin{document}
\begin{titlepage} 
\begin{flushright}
CERN-TH-2016-213\\
DCPT-16/39\\
CCTP-2016-15\\
ITCP-IPP-2016/10
\vskip -1cm
\end{flushright}
\vskip 1.0cm
\begin{center}
\font\titlerm=cmr10 scaled\magstep4
    \font\titlei=cmmi10 scaled\magstep4
    \font\titleis=cmmi7 scaled\magstep4
    \centerline{\LARGE \titlerm 
    Large-$N$ correlation functions}
    \vskip 0.3cm
    \centerline{\LARGE \titlerm
    in $\NN=2$ superconformal QCD}
\vskip 1cm
{Marco Baggio,$^a$ Vasilis Niarchos,$^b$ Kyriakos Papadodimas,$^{c,d}$ Gideon Vos$^e$ }\\
\vskip 0.5cm
{\it $^a$Institute for Theoretical Physics, KU Leuven, 3001 Leuven, Belgium}\\

\medskip
{\it $^b$Department of Mathematical Sciences and Center for Particle Theory}\\
{\it Durham University, Durham, DH1 3LE, UK}\\
\medskip
{\it $^{c}$Theory Group, Physics Department, CERN, CH-1211 Geneva 23, Switzerland}\\
{\it $^{d,e}$Van Swinderen Institute for Particle Physics and Gravity, University of Groningen, Nijenborgh 4,
9747 AG Groningen, The Netherlands}\\
\medskip
{$^a$marco.baggio@kuleuven.be, $^b$vasileios.niarchos@durham.ac.uk}\\
{$^{c,d}$kyriakos.papadodimas@cern.ch, $^e$g.vos@rug.nl}

\end{center}
\vskip .5cm
\centerline{\bf Abstract}

\baselineskip 20pt
%

\vskip .5cm 
\noindent
We study extremal correlation functions of chiral primary operators in the large-$N$ $SU(N)$ $\NN=2$ superconformal QCD
theory and present new results based on supersymmetric localization. We discuss extensively the basis-independent data
that can be extracted from these correlators using the leading order large-$N$ matrix model free energy given by the
four-sphere partition function. Special emphasis is given to single-trace 2- and 3-point functions as well as a new
class of observables that are scalars on the conformal manifold. These new observables are particular quadratic
combinations of the structure constants of the chiral ring. At weak 't Hooft coupling we present perturbative results
that, in principle, can be extended to arbitrarily high order. We obtain closed-form expressions up to the first
subleading order. At strong coupling we provide analogous results based on an approximate Wiener-Hopf method.

\vfill
\noindent
\end{titlepage}\vfill\eject

\hypersetup{pageanchor=true}

\setcounter{equation}{0}

\pagestyle{empty}
\small
\vspace*{-0.7cm}
{
\hypersetup{linkcolor=black}
\tableofcontents
}
\normalsize

\pagestyle{plain}
\setcounter{page}{1}
 
 \newpage
 
\rightline{
{\it Dedicated to the memory of Ioannis Bakas}}

\section{Introduction}
\label{setup}

References \cite{Baggio:2014ioa, Baggio:2014sna} computed the exact (extremal) correlation functions of 
$\NN=2$ chiral primary operators in the 4d ${\cal N}=2$ superconformal gauge theory with $SU(2)$ gauge group 
coupled to $N_f=4$ massless hypermultiplets. These correlation functions are highly nontrivial functions of the 
complexified coupling
constant $\tau={\theta \over 2\pi } + i{4\pi \over g^2}$ and include all-order perturbative and instanton corrections.
At the moment, they are the only known example of nontrivial, exactly computed 3-point functions in a 4d QFT. The
computation of \cite{Baggio:2014ioa, Baggio:2014sna} relied on the constraints imposed on the chiral ring correlators by
the 4d $tt^*$ equations \cite{Papadodimas:2009eu}, together with input from supersymmetric localization
\cite{Pestun:2007rz}, and made use of the relation, proposed in \cite{Gerchkovitz:2014gta}, between the sphere partition
function $\ZZ_{{S}^4}$ and the Zamolodchikov metric on the conformal manifold. The relationship between 
extremal correlation
functions in the $\NN=2$ chiral ring and the sphere partition function was further clarified and extended in
\cite{Gerchkovitz:2016gxx}, which paved the way towards concrete computations in general 4d $\NN=2$ SCFTs with 
conformal manifolds. More generally, it would be interesting to know if there are also other correlation
functions that can be computed in practice by employing similar techniques (see \cite{Dedushenko:2016jxl}
for a recent analogous computation of correlation functions in 3d $\NN=4$ superconformal field theories). 

In this paper we consider the family of ${\cal N}=2$ superconformal field theories with gauge group $SU(N)$ and 
$N_f=2N$ hypermultiplets. We focus on the large-$N$ 't Hooft-Veneziano limit and explain how correlators of chiral 
primary operators can be computed as a function of the 't Hooft coupling $\lambda = g^2 N$. One reason why these 
correlators are interesting is that they encode information about a putative string theory dual for this family of 
large-$N$ theories\footnote{Since we consider a 't Hooft-Veneziano limit where 
the ratio $N_f/N=2$ is fixed and non-vanishing, this duality would have the peculiar feature where 
mesonic hypermultiplet bilinears would lead to an $O(N^2)$ number of gauge invariant operators with 
low conformal dimension. A related discussion of similar limits in two-dimensional theories can be found in 
\cite{Kiritsis:2010xc}. We thank S. Minwalla for comments related to this feature.} 
(see \cite{Gadde:2009dj} for an earlier discussion of such duality). Moreover, similar techniques 
could be applied to closely related theories (e.g.\ $\NN=2$ orbifolds of $\NN=4$ super-Yang-Mills theory (SYM) 
\cite{Kachru:1998ys}) with known AdS/CFT duals, and lessons obtained in this paper could be easily extended there 
as well. A recent discussion of conformal manifolds in the context of the AdS/CFT correspondence from the 
supergravity point of view can be found in \cite{Louis:2015dca}.
More generally, having a solid understanding of a large-$N$ correlator as an exact function of $\lambda$ in 
QFT, at leading and subleading orders in the $1/N$-expansion, could be a useful guide towards a concrete analysis 
of various formal aspects of the large-$N$ expansion in a full-fledged 4d gauge theory.

$\NN=2$ chiral primary correlators in the 't Hooft limit of the ${\cal N}=2$ $SU(N)\,,\,N_f=2N$ theory were recently
considered in \cite{Rodriguez-Gomez:2016ijh,Rodriguez-Gomez:2016cem}, 
where 2-point functions of single-trace chiral primaries were computed 
perturbatively in $\lambda$ at leading order in $1/N$. In this paper we substantially extend these results by computing
more general correlation functions at large $N$. Specifically, we focus on two main classes of observables. 

The first are 3-point functions of chiral primary single-trace operators. 3-point functions of chiral primary operators
in $\NN=4$ SYM theory have of course been widely studied in the context of holography starting with
\cite{Lee:1998bxa,D'Hoker:1998tz}. In the ${\cal N} = 2$ theory, all the non-vanishing 3-point functions are extremal,
and are especially sensitive to mixing with multi-trace operators \cite{D'Hoker:1999ea}. We point out that there is a
well-motivated and unambiguous definition of the basis of chiral primary operators near the weak-coupling point based on
parallel transport that is formulated in terms of a natural connection on the space of operators in conformal
perturbation theory. This definition works particularly well in our class of theories in the large-$N$ limit, where the
conformal manifold is essentially one-dimensional.
A different basis of chiral primary operators is defined implicitly through the relation with the $S^4$ partition
function $\ZZ_{S^4}$ \cite{Gerchkovitz:2016gxx}. We compute 3-point functions of the form $\langle {\rm
Tr}\varphi^{k_1}\, {\rm Tr}\varphi^{k_2}\, {\rm Tr}\overline\varphi^{k_1+k_2}\rangle$ in the first few orders in
$\lambda$ around the weak coupling point in both bases. We check that to leading order our methods reproduce the results
of \cite{Lee:1998bxa} for the ${\cal N}=4$ SYM, as expected. Unlike the ${\cal N}=4$ SYM theory, however, in ${\cal
N}=2$ theories correlators of chiral primaries receive quantum corrections that we can easily compute up to any desired
order in $\lambda$.

In order to bypass the subtleties that arise from the mixing between single-trace and multi-trace operators
we also consider a new class of observables obtained by certain quadratic combinations of
the chiral ring structure constants. These quantities are geometric scalars on the conformal manifold, they are
manifestly independent of the choice of basis, and therefore can be meaningfully computed and compared
at arbitrary values of the coupling constant. Furthermore, an infinite subset of them obey a
very simple recursion relation, coming from the \tts equations, that can be solved explicitly in terms of 
2-point function data immediately available at large $N$.

We show how both classes of observables can be computed at leading order in the large-$N$ limit from the 
planar free energy $F_0$ of the theory on $S^4$ deformed by higher chiral primary sources. The
latter is also the planar free energy of a corresponding matrix model, which arises from localization, and can be
determined from the solution of the saddle-point integral equation
\be
\label{integraleq}
\int_{\mu_-}^{\mu_+} dx  \left[{1\over x-y} - K(x-y)\right] \rho(y) = {8 \pi^2 \over \lambda}x - K(x) + \sum_{n=2}^M t_n x^n
\ee where $K(x) = 2x \sum_{n=1}^\infty \left( \frac{1}{n} - \frac{n}{n^2+x^2} \right)$. The sum on the r.h.s.\ originates from
the higher chiral primary source deformations of the theory. It is a polynomial whose degree is suitably adjusted to the
correlator we want to compute. The planar free energy $F_0$ follows directly from the eigenvalue density $\rho(x)$. Eq.\
\eqref{integraleq} was first considered in \cite{Passerini:2011fe} and further used in \cite{Rodriguez-Gomez:2016ijh}.
We have not been able to solve \eqref{integraleq} analytically for arbitrary values of $\lambda$, so we will limit
ourselves to analyzing its solutions in two regimes, at weak and strong coupling $\lambda$.
Given a solution of \eqref{integraleq} (approximate or exact), there is a well-defined procedure
\cite{Gerchkovitz:2016gxx,Baggio:2014ioa} to recover correlation functions of the physical theory by combining
appropriate derivatives of $F_0$.

Computations based on the weak coupling expansion of the solutions of \eqref{integraleq} are pretty straightforward and,
technically, they follow closely the logic of \cite{Passerini:2011fe,Rodriguez-Gomez:2016ijh}. At strong coupling
the analysis of eq.\ \eqref{integraleq} is considerably harder. As was first pointed out in \cite{Passerini:2011fe}
approximate solutions can be obtained with the use of the Wiener-Hopf method.\footnote{We should point out that these approximations are not parametrically controlled, so we cannot prove conclusively that the large-$\lambda$ scalings
obtained in this way persist in the exact solution of the saddle-point equations.} Using this method we estimate the
large-$\lambda$ scaling of 2-point functions of single-trace operators in the chiral ring, extending partial results in
\cite{Rodriguez-Gomez:2016ijh}, and the large-$\lambda$ scaling of 3-point functions. The large-$\lambda$ scaling of
2-point functions is also discussed from an independent point of view based on the analysis of the density of connected
2-point functions in the matrix model.

\vspace{0.5cm}
\noindent
{\bf Plan of the paper and summary of the main results.} 
In the main text of the paper we focus on properties and results of correlators on
$\IR^4$. Intermediate results based on localization and the corresponding matrix model are relegated to the appendices,
where the reader can find all the pertinent details.

In section \ref{generics} we discuss in detail the correlation functions of interest and we set the conventions that are
used in the rest of the paper. In addition, we review the relation between extremal correlation functions in the $\NN=2$
chiral ring, the deformed partition function on $S^4$ and the matrix model that arises from localization.

In section \ref{largeN} we discuss general properties of correlation functions in the large-$N$ limit. We explain what
contributions can be extracted from the leading order large-$N$ free energy of the $S^4$ partition function and how
issues involving the mixing of single-trace and multi-trace operators affect our computations. We also define
appropriate quadratic combinations of the structure constants and show that they obey a recursion relation, coming from
the \tts equations, that can be solved in closed form.

Results specific to the weak coupling expansion of the theory are presented in section \ref{weakcoupling}. We provide
closed form expressions for 2- and 3-point functions both at tree level and at the first nontrivial subleading order in
perturbation theory. Along the way, we present a method, specific to the large-$N$ limit, that allows us to determine
the correlation functions of single-trace operators without going through the full Gram-Schmidt orthogonalization
procedure proposed in \cite{Gerchkovitz:2016gxx}. In this section we also discuss how the use of parallel transport on
the conformal manifold leads to unambiguous perturbative expressions for the single-trace 3-point functions.
The basis-independent structure constant squared combinations, defined in section \ref{largeN}, are computed 
perturbatively in $\lambda$ at the end of the section.

Finally, partial results in the strong coupling limit of the theory are discussed in section \ref{strong}. We emphasize
the large-$\lambda$ scaling of 2- and 3-point functions and discuss the technical difficulties associated to the current use of
the Wiener-Hopf method.

Four appendices at the end of the paper provide the technical background for the computations presented in the main
text. Appendix \ref{deformedmatrix} summarizes the matrix model that arises from localization and the corresponding
saddle-point equations in the large-$N$ limit. In this appendix the reader can also find the derivation of an integral 
equation obeyed by the density of connected 2-point functions, as well as an explicit solution of this equation at 
infinite 't Hooft coupling. Appendix \ref{appweak} describes the perturbative solution of the
saddle-point equations at weak coupling and appendix \ref{appstrong} the approximate solution based on the Wiener-
Hopf method. Appendix \ref{proof} provides the proof of a technically efficient general relation between 3-point functions 
in the gauge theory and derivatives of the matrix model planar free energy in the large-$N$ limit.

\section{Exact 2- and 3-point functions in the $\NN=2$ chiral ring}
\label{generics}

In this paper we focus on {\it extremal} correlation functions of $\NN=2$ chiral primary operators 
in a specific class of $\NN=2$ superconformal field theories defined as $\NN=2$ SYM theory with 
gauge group $SU(N)$ coupled to $2N$ hypermultiplets (in short, $SU(N)$ $\NN=2$ superconformal-QCD, or 
$\NN=2$ SCQCD). 
We will mostly follow the conventions of \cite{Baggio:2014ioa,Baggio:2015vxa}, where one can also find
a detailed description of generic properties of the $\NN=2$ chiral primary operators and further useful references to 
the literature. 

We begin with a quick summary of the operators of interest tailored 
to the specific features of the $\NN=2$ SCQCD theories and the goals of this paper. 
Then, we proceed to define the correlation functions that will play a central role
in our discussion and to summarize recent developments that allow their exact non-perturbative computation. 
Along the way, we emphasize the implications of the recent developments on 2- and 3-point functions.

\vspace{0.3cm}
\noindent
{\bf Operator notation.} In the course of the paper we will consider the $\NN=2$ theory either on $\IR^4$ 
or $S^4$. To keep the distinction between these cases explicit at all times,
we will refer to the chiral primary operators on $\IR^4$ as $O_K$ and the corresponding operators 
on $S^4$ as $O_K^{S^4}$. $K$ is 
an appropriate multi-index that labels the operator. Moreover, for notational economy we will frequently refer to
single-trace generators $\Tr[\varphi^k]$ on $\IR^4$ as $k$ 
inside correlation functions, double-trace operators $\Tr[\varphi^{k_1}]\Tr[\varphi^{k_2}]$ as $k_1k_2$, etc.  
The $S^4$ operators may acquire a further label, $O^{\IR^4}$, 
that refers to specific linear combinations of single/multi-trace operators to be defined.

\vspace{0.3cm}
\noindent
{\bf Correlation function notation.}
Correlation functions on $\IR^4$ will be denoted as $\langle \cdots \rangle_{\IR^4}$ (or simply as 
$\langle \cdots\rangle$ without index), correlation functions on $S^4$ as $\langle \cdots \rangle_{S^4}$ 
and correlation functions on the associated matrix model as $\llangle \cdots \rrangle$.

\subsection{$SU(N)$ $\NN=2$ chiral ring}
\label{generics1}

We begin by considering the $SU(N)$ $\NN=2$ SCQCD theory on flat space, ${\mathbb R}^4$.
The $\NN=2$ chiral primary operators are, by definition, local superconformal primary operators annihilated by 
all four left-chiral Poincar\'e supercharges $\barM Q^i_{\dot\alpha}$, where $i=1,2$ is an $SU(2)_R$ index and
$\dot \alpha=\pm$ a spinor index. 
In the $\NN=2$ SCQCD theory these operators have a simple description as 
generic multi-trace operators of the adjoint complex scalar field $\varphi$ in the $\NN=2$ vector multiplet. Using 
a multi-index $K=\{ n_\ell \}$ we denote them as
\beq\label{gen1aa}
O_K \equiv O_{\{ n_\ell\}} \propto \prod_{\ell=1}^{N-1} \left( \Tr \left[ \varphi^{\ell+1} \right] \right)^{n_\ell}
~,
\eeq
where $n_\ell$ are arbitrary non-negative integers.
The proportionality symbol refers to an overall normalization factor that will be fixed later. Corresponding
multi-trace operators built out of the complex conjugate field $\barM \varphi$ will be denoted as $\barM O_K$; those
are $\NN=2$ anti-chiral primary operators annihilated by all four right-chiral Poincar\'e supercharges
$Q^i_\alpha$. 

The scaling dimension $\Delta_K$ of each of the operators $O_K$ is half their $U(1)_R$ charge $R_K$
\beq\label{gen1ab}
\Delta_K = \frac{R_K}{2} = \sum_{\ell=1}^{N-1} (\ell+1) n_\ell
~.\eeq
This relation holds non-perturbatively for generic values of the exactly marginal coupling constant of the theory 
$\tau = \frac{\theta}{2\pi} + i \frac{4\pi}{g^2}$, where as usual $\theta$ is the theta-angle of the theory and 
$g$ the gauge coupling.

It is clear from the definition \eqref{gen1aa} that the full class of chiral primary multi-trace operators $O_K$ can be
generated by Operator Product Expansion (OPE) multiplication from a finite set of $N-1$ single-trace operators 
$\Tr\left[ \varphi^{\ell+1} \right]$, 
$\ell=1,\ldots,N-1$. 
In what follows we will adopt a normalization convention, consistent with the so-called 
holomorphic gauge \cite{Baggio:2014ioa}, where the leading term in the OPE
between two chiral primary operators,
\beq
O_K(x) O_L(0) = C_{KL}^M O_M(0)+\ldots~,
\eeq
is
\beq\label{gen1ac}
O_K(x) O_L(0) = O_{K+L}(0)+\ldots
~.\eeq
The dots indicate higher-dimension descendant operators.
$O_{K+L}$ is the multi-trace operator $:O_K O_L:$. The absence of a spacetime singularity in the OPE of
two chiral primary operators is a characteristic property of chiral primary operators.
The convention \eqref{gen1ac}, which sets\footnote{In this expression $\delta_I^J$ is 
the obvious multi-index Kronecker delta.} 
\beq\label{gen1ad}
C_{KL}^M= \delta^M_{K+L}
~,\eeq 
allows us to fix the 
normalization of all multi-trace operators $O_K$ in terms of the normalization of the single-trace generators 
$\Tr \left[ \varphi^{\ell+1} \right]$.

\subsection{Extremal correlation functions in the $\NN=2$ chiral ring}

The main interest of the paper lies in the so-called {\it extremal} correlation functions, 
defined as correlation functions of chiral and anti-chiral primary operators with a single anti-chiral insertion
\beq\label{gen2aa}
\langle O_{K_1}(x_1) O_{K_2}(x_2) \cdots \barM O_{K_n}(x_n) \rangle
~.\eeq
The $U(1)_R$ charge conservation requires the $R$-charge relation
\beq\label{gen2ab}
\sum_{i=1}^{n-1} R_{K_i}= - R_{K_n}
~,\eeq
otherwise the correlator vanishes.

In \cite{Baggio:2014sna,Baggio:2014ioa} it was argued that all extremal 
correlation functions can be reduced to the computation of the 2- and 3-point functions, respectively
\beq\label{gen2aca}
\langle O_K(x_1) \barM O_L(x_2) \rangle = \frac{g_{K\bar L}}{|x_{12}|^{2\Delta}}~, ~~ x_{12}\equiv x_1-x_2
~,
\eeq
\beq\label{gen2acb}
\langle O_{K}(x_1) O_{L}(x_2) \barM O_{M}(x_3) \rangle = 
\frac{C_{KL\barM M}}{|x_{12}|^{\Delta_K+\Delta_L-\Delta_M} |x_{13}|^{\Delta_K+\Delta_M-\Delta_L}
|x_{23}|^{\Delta_L+\Delta_M-\Delta_K}}
~.\eeq  
$\Delta$ is the common scaling dimension of the two insertions in the 2-point function \eqref{gen2aca}, and 
$\Delta_K$ etc.\ the scaling dimensions of each operator in the 3-point function \eqref{gen2acb}. The interesting
datum in each of these correlation functions is the position independent, but generally coupling constant dependent,
numerator $g_{K\bar L}$ in the 2-point functions and $C_{KL\barM M}$ in the 3-point functions. In the rest of the 
text it will be convenient to refer to these coefficients using the notation 
\beq\label{gen2acc}
\langle O_K,\barM O_L\rangle \equiv g_{K\barM L}~, ~~
\langle O_K, O_L, \barM O_M\rangle \equiv C_{KL\barM M}
~.\eeq

There is a simple well-known relation between the 2- and 3-point function coefficients and the OPE coefficients
$C_{KL}^M$ in the $\NN=2$ chiral ring
\beq\label{gen2ad}
C_{KL\barM M} = C_{KL}^I \, g_{I\barM M}
~.\eeq
Notice that by using the convention
\eqref{gen1ad} equation \eqref{gen2ad} reduces to  
\beq\label{gen2ae}
C_{KL\barM M} = g_{K+L, \barM M} 
~.\eeq

As an explicit illustration of this relation, consider the computation of the 3-point function of
single-trace operators 
\beq
\langle k_1, k_2, \overline{k_1+k_2} \rangle
= \left\langle \Tr\left[\varphi^{k_1}\right], \Tr\left[\varphi^{k_2}\right], \Tr\left[\barM \varphi^{k_1+k_2}\right]\right\rangle
~,
\eeq 
where following the aforementioned convention we denote the single trace operator $\Tr[\varphi^k]$ simply as $k$ in 
a correlation function.
Equation \eqref{gen2ae} implies that this is equal to 
\beq
\langle k_1, k_2, \overline{k_1+k_2} \rangle
= \langle k_1 k_2, \overline{k_1+k_2} \rangle
=\left\langle \left(  \Tr\left[\varphi^{k_1}\right]  \Tr\left[\varphi^{k_2}\right] \right),
\Tr\left[\barM \varphi^{k_1+k_2}\right] \right\rangle
~,
\eeq
which is a 2-point function between a double-trace operator and a single-trace operator.

\subsection{2- and 3-point functions from $S^4$ partition functions and matrix models}
\label{S4correspondence}

So far we exclusively discussed correlation functions of the $\NN=2$ SCQCD theory on ${\mathbb R}^4$.
In recent developments, however, a concrete relation has been put forward between the 
2-point function coefficients $\langle O_K,\barM O_M\rangle$ of the theory on $\IR^4$ and the derivatives of 
a suitably deformed partition function of the theory on the four-sphere $S^4$ 
\cite{Gerchkovitz:2014gta,Gomis:2014woa,Gomis:2015yaa,Gerchkovitz:2016gxx}. 
The latter is further related by supersymmetric localization \cite{Pestun:2007rz} to the partition function of 
a corresponding matrix model. 

Let us briefly review the main elements of this relation and set up the 
appropriate notation. For additional explanations and details we refer the reader to the original work in 
\cite{Gerchkovitz:2014gta,Gomis:2014woa,Gomis:2015yaa,Gerchkovitz:2016gxx}.

\subsubsection{Deformed partition functions on $S^4$ and their localization}

The first step of the procedure starts, quite generally, by placing the $\NN=2$ 
superconformal field theory on $S^4$ in a manner that preserves the supergroup of a general 
massive theory, $osp(2|4)$. In addition, we deform the theory by F-term interactions
that are upper components of short multiplets containing the $\NN=2$ chiral primary fields $O_K$. 
It is enough for our purposes 
to consider deformations restricted to the single-trace chiral primary fields $\Tr\left[ \varphi^k \right]$. In 
superspace form the deformations of interest are 
\beq\label{gen3aa}
\delta \SS = -\frac{1}{32\pi^2} \sum_{n=2}^N \int d^4 x \int d^4\theta\, \EE \tau_n \Tr[ \varphi^n] +{\rm c.c.}
~,\eeq
where $\EE$ is the $\NN=2$ chiral density. $\tau_2\equiv \tau= \frac{\theta}{2\pi}+i \frac{4\pi}{g^2}$ is the 
exactly marginal deformation of the $\NN=2$ SCQCD theory. 

Now consider the partition function of this theory
\beq
\ZZ_{S^4}\left( \tau_n, \bar \tau_n \right)~.
\eeq
The finite part of this quantity is physical \cite{Gerchkovitz:2014gta,Gomis:2015yaa} 
and depends non-trivially on the complex couplings $(\tau_n, \bar\tau_n)$.
Interestingly, although this quantity is given by a complicated path integral, it can be reduced by 
supersymmetric localization to a corresponding matrix integral that can be analyzed with standard methods
\cite{Pestun:2007rz}. 
The precise form of the matrix integral in the case of the $\NN=2$ SCQCD theories is presented in appendix 
\ref{deformedmatrix}.

\subsubsection{Relation between the $S^4$ partition function and 2-point functions on $\IR^4$}
\label{fromStoR}

Recently, \cite{Gerchkovitz:2016gxx} put forward a concrete general prescription 
that relates the $\tau_n$-derivatives of $\ZZ_{S^4}$ to 
the flat-space 2-point function coefficients $\langle O_K, \barM O_M\rangle$. 
One way to summarize the prescription is the following. 

Assume we want to evaluate the 2-point function coefficient 
$\langle O_K, \barM O_M\rangle_{\IR^4}$\footnote{At this point
we will include an index $\IR^4$ or $S^4$ in the notation of the correlation functions to denote explicitly whether we
refer to correlation function on $\IR^4$ or $S^4$.} for two operators $O_K$, $\barM O_M$ of the same 
scaling dimension $\Delta$ in the $\NN=2$ chiral ring. Consider the same (single or multi-trace) operators on $S^4$ 
---$O_K^{S^4}$ for the counterpart of $O_K$---
and construct linear combinations $O_K^{\IR^4}$ where operators at scaling dimension 
$\Delta$ mix with all operators of smaller dimension $\Delta-2, \Delta-4,\ldots$ (including the identity operator when 
$\Delta$ is even)
\beq\label{gen4aa}
O_K^{\IR^4} = O_K^{S^4} + \sum_{I\in S_\Delta} a_I O^{S^4}_I~, ~~
\barM O_M^{\IR^4} = \barM O_M^{S^4} + \sum_{I\in S_\Delta} \barM b_I O^{S^4}_I~.
\eeq
The sum $\sum_I$ runs over the set $S_\Delta$, which is defined to include all the chiral primaries 
of scaling dimension $\Delta_I < \Delta$, 
$\Delta_I= \Delta$ mod 2. The coefficients $a_I$, $\barM b_I$ are clearly dimensionful and therefore proportional to 
an appropriate power of the sphere radius. They are fully fixed by implementing the Gram-Schmidt orthogonalization 
procedure,
\beq\label{gen4ab}
\langle O_K^{\IR^4} , \barM O_L^{S^4} \rangle_{S^4} = 0~, ~~
\langle O_L^{\IR^4} , \barM O_M^{S^4} \rangle_{S^4} = 0
~~{\rm for~all}~~ L \in S_\Delta
~.
\eeq
The key statement of \cite{Gerchkovitz:2016gxx} is the relation\footnote{Notice that the conformal mapping between
the sphere and the plane introduces an additional factor of $4^\Delta$ in the relation between the sphere and the plane
2-point functions. To avoid clutter in the equations, we absorb this factor in the normalization of the operators $O_K$.
Of course this has no effect on the normalized correlators that we discuss in the rest of the paper.}
\beq\label{gen4ac}
\langle O_K, \barM O_M \rangle_{\IR^4} 
= \langle O_K^{\IR^4} , O_M^{\IR^4} \rangle_{S^4}
~.\eeq
Employing eqs.\ \eqref{gen4aa}, \eqref{gen4ab} we obtain
\beq\label{gen4ad}
\langle O_K, \barM O_M \rangle_{\IR^4} 
= \langle O_K^{S^4}, \barM O_M^{S^4} \rangle_{S^4} 
- \sum_{I,J\in S_\Delta} \langle O_K^{S^4}, \barM O_I^{S^4} \rangle_{S^4} \left( \A^{-1} \right)_{IJ}
\langle O_J^{S^4}, \barM O_M^{S^4} \rangle_{S^4}
~,\eeq
where the matrix $\A$ has, by definition, the elements
\beq\label{gen4ae}
\A_{IJ} = \langle O_I^{S^4}, \barM O_J^{S^4} \rangle_{S^4}~, ~~ I, J \in S_\Delta
~.
\eeq
In eq.\ \eqref{gen4ad} we assumed that the matrix $\A$ is invertible, which is a prerequisite for the prescription of
\cite{Gerchkovitz:2016gxx} to work properly.

The final element is the statement that the 2-point function coefficients 
$\langle O_I^{S^4}, \barM O_J^{S^4} \rangle_{S^4}$ are simply given by derivatives of the deformed 
$S^4$ partition function as follows
\bea\label{gen4af}
\langle O_{\{k_\ell \}}^{S^4}, \barM O_{\{ m_\ell' \}}^{S^4} \rangle_{S^4} 
&=& \frac{1}{\ZZ_{S^4}} \prod_{\ell,\ell'=1}^{N-1} \frac{\d}{\d \tau_{\ell+1}} \frac{\d}{\d \bar \tau_{\ell'+1}} \ZZ_{S^4}
\bigg |_{\tau_2=\tau,\, \tau_k=0,\, k\neq 2}
\cr
&\equiv& \prod_{\ell,\ell'=1}^{N-1} \llangle \left( \Tr \left[ \varphi^{\ell+1} \right] \right)^{n_\ell}
\left( \Tr \left[ \barM \varphi^{\ell'+1} \right] \right)^{n_{\ell'}} \rrangle
~.\eea
$\llangle \cdots \rrangle$ denotes a correlation function in the matrix model of appendix \ref{deformedmatrix}.

Having determined the 2-point functions in this manner we have essentially fixed the normalization conventions for
all the $\NN=2$ chiral primary operators. At this point one should wonder if this prescription is consistent with 
the choice \eqref{gen1ad} for the OPE coefficients. Following the work in \cite{Baggio:2014ioa}, Ref.\ 
\cite{Gerchkovitz:2016gxx} demonstrated that the ansatz \eqref{gen4af} satisfies the full set of \tts equations with 
\eqref{gen1ad} incorporated. This is a strong explicit check that \eqref{gen4af} is indeed consistent with \eqref{gen1ad}.

\subsubsection{Formulae for 3-point functions}
\label{3general}

Combining equations \eqref{gen2ae}, \eqref{gen4ad}, \eqref{gen4af} we are now in position to write down an explicit
formula for 3-point functions on $\IR^4$
\beq\label{gen5aa}
\langle O_K, O_L, \barM O_M \rangle_{\IR^4} = 
\langle O_{K+L}^{S^4}, \barM O_M^{S^4} \rangle_{S^4} 
- \sum_{I,J\in S_{\Delta_M}} \langle O_{K+L}^{S^4}, \barM O_I^{S^4} \rangle_{S^4} \left( \A^{-1} \right)_{IJ}
\langle O_J^{S^4}, \barM O_M^{S^4} \rangle_{S^4}
~.
\eeq
All 2-point functions on the r.h.s.\ of this equation can be expressed via \eqref{gen4af} in terms of derivatives of the 
deformed $S^4$ partition function, or alternatively in terms of derivatives of the free energy 
\beq\label{gen5ab}
F= -\log \ZZ_{S^4}
\eeq
of the corresponding matrix model.

As an explicit example consider again the 3-point function of three single-trace operators. The above prescription gives 
\bea\label{gen5ac}
&&\left\langle \Tr\left[\varphi^{k_1}\right], \Tr\left[\varphi^{k_2}\right], \Tr\left[\barM \varphi^{k_1+k_2}\right]\right\rangle_{\IR^4}
= \llangle \Tr\left[\varphi^{k_1}\right]\Tr\left[\varphi^{k_2}\right] \Tr\left[\bar\varphi^{k_1+k_2}\right] \rrangle
\cr
&&- \sum_{I,J\in S_{k_1+k_2}} 
\llangle \Tr\left[\varphi^{k_1}\right]\Tr\left[\varphi^{k_2}\right] \barM O_I^{S^4} \rrangle
\left( \A^{-1} \right)_{IJ}
\llangle O_J^{S^4}  \Tr\left[\bar\varphi^{k_1+k_2}\right] \rrangle
~.\eea
Obviously, the structure of the sum on the r.h.s becomes increasingly complicated with increasing scaling dimension.

For a more concrete illustration consider a 3-point function that involves the lowest lying single-trace operators, e.g.\
$\left\langle \Tr\left[\varphi^{2}\right], \Tr\left[\varphi^{2}\right], \Tr\left[\barM \varphi^{4}\right]\right\rangle_{\IR^4}$.
In this case the matrix $\A$ appearing on the r.h.s.\ of eq.\ \eqref{gen5ac} is
\beq\label{gen5ada}
\A= \left(
\begin{array}{c c}
\llangle \Tr\left[\varphi^2\right] \Tr\left[ \bar\varphi^2 \right]\rrangle &~  \llangle \Tr\left[\varphi^2\right] \rrangle \\
\llangle \Tr\left[ \bar\varphi^2 \right]\rrangle &~ 1 \\
\end{array}
\right)
=
\left(
\begin{array}{c c}
-\d_{\tau_2}\d_{\bar \tau_2} F + \d_{\tau_2} F \, \d_{\bar\tau_2} F &~ - \d_{\tau_2} F \\
-\d_{\bar\tau_2} F &~ 1 \\
\end{array}
\right)
~.\eeq
Explicit evaluation gives the following simple 2- and 3-point function formulae
\beq\label{gen5ae}
\left\langle \Tr\left[\varphi^2 \right],  \Tr\left[\bar\varphi^2 \right] \right\rangle_{\IR^4} 
=- \d_{\tau_2} \d_{\bar\tau_2} F~, ~~
\eeq
\beq\label{gen5af}
\left\langle \Tr\left[\varphi^4 \right],  \Tr\left[\bar\varphi^4 \right] \right\rangle_{\IR^4} 
=- \d_{\tau_4} \d_{\bar\tau_4} F + \frac{\d_{\tau_2} \d_{\bar\tau_4} F \, \d_{\tau_4} \d_{\bar\tau_2} F}
{\d_{\tau_2}\d_{\bar\tau_2}F}
~,
\eeq
\beq\label{gen5ad}
\left\langle \Tr\left[\varphi^{2}\right], \Tr\left[\varphi^{2}\right], \Tr\left[\barM \varphi^{4}\right]\right\rangle_{\IR^4}
= - \d^2_{\tau_2} \d_{\bar \tau_4} F 
+ \frac{\d_{\tau_2}\d_{\bar\tau_4} F\, \d^2_{\tau_2}\d_{\bar \tau_2} F}{\d_{\tau_2}\d_{\bar \tau_2}F} 
~,
\eeq
where the final result is expressed directly in terms of derivatives of the matrix model free energy $F$.

The correlation functions of operators with higher scaling dimensions can be expressed similarly 
solely in terms of $F$, but the final expression is considerably more complicated. 
Further simplifications occur, however, in the large-$N$ limit, which is the main topic of the following sections.

\section{Correlation functions at large $N$}
\label{largeN}

In this section we study extremal correlation functions in the large-$N$ limit and their
relation to the matrix model. Due to large-$N$ factorization, the behavior of
correlators involving multi-trace operators is dominated at large $N$ by the factorized answer. Therefore, we
introduce a notion of ``connected'' 2-point functions, which involves, as usual, the full 2-point functions minus the
factorized pieces. We argue that these correlators can be determined
by the leading contribution to the free energy in the large-$N$ limit, which in turn can be computed by 
the saddle-point method.

At a later part of this section we specialize to the two main classes of observables that we are interested in. First, we
study in detail the relation between single-trace 3-point functions and the free energy, and discuss some useful
simplifications that occur at large $N$. We also discuss in detail issues related to mixing between
single- and multi-trace operators, which in principle can affect the results of our computation, and propose one way to
get around these difficulties by using the natural connection provided by conformal perturbation theory.

Lastly, we define a new interesting class of observables, which are quadratic combinations of the structure constants
that enjoy many useful properties. Most notably, these observables are manifestly free from
ambiguities related to the choice of basis of chiral operators, so they are not affected by the
subtleties associated to large-$N$ mixing between single- and multi-trace operators. In addition, the \tts
equations provide a very simple recursion relation for these observables, which can be solved in closed 
form in terms of simple geometric data on the conformal manifold.

The explicit analysis of these quantities at weak and strong coupling is the subject of subsequent sections.

\subsection{Correlation functions and the matrix model free energy at large $N$}

We consider the large-$N$ limit at fixed 't Hooft coupling constant $\lambda = g^2 N$. Similarly, we rescale the sources
of the higher Casimir operators so that the parameters\footnote{In \cite{Rodriguez-Gomez:2016ijh}, a different
convention for the couplings was used, namely $g^{\mathrm{there}}_{n} = \pi^{\frac{n}{2}} g^{\mathrm{here}}_n$. We find
that our choice is more convenient for the purpose of this paper, as it avoids various explicit factors of $\pi$ that
would otherwise appear in intermediate formulae. This effectively corresponds to a different overall normalization of
the chiral operators compared to \cite{Rodriguez-Gomez:2016ijh}, which of course does not have any effect on the
normalized correlators.}
\beq
\label{scaledcouplings}
g_{n} =  \frac{2}{N} \mathrm{Im}\,\tau_n~, ~~ n=2,3,\ldots
\eeq
are kept fixed in the limit, as was done in \cite{Rodriguez-Gomez:2016ijh}. The free energy \eqref{gen5ab} has the
following large-$N$ expansion
\beq
\label{eqn:fee}
F = N^2 F_0(\{g_n\}) + F_1(\{g_n\}) + \ldots~
\eeq
$F_0$ is the leading large-$N$ contribution. It can be evaluated using the saddle-point approximation, 
details of which we review in appendix \ref{deformedmatrix}. 

In the previous section, we reviewed how generic 2-point functions (of single-trace or multi-trace operators) in the
chiral ring on $\IR^4$ can be expressed in terms of an algebraic functional of derivatives of the free energy $F$. In
the large-$N$ limit, and after the Gram-Schmidt procedure has been properly applied, the result contains a finite number
of derivatives of $F$ with respect to the parameters $g_n$. The leading contribution to this result comes from $F_0$,
and may scale with $N$ in different ways depending on the specifics of the operator insertions.

For instance, the 2-point function of two single-trace operators
\beq
\label{11a}
\left<k \, , \barM k\right> \sim \OO(N^0)~,
\eeq
scales at large $N$ as a constant. Similarly, the 2-point function of a multi-trace operator 
with a single-trace operator scales like
\beq
\left<k_1\cdots k_m\, ,\barM k\right> \sim \OO(N^{-m+1})~.
\eeq
So, for example, the leading order scaling of the 2-point function of a double-trace and a single-trace operator
is of order $\OO(N^{-1})$ in agreement with the 3-point function scaling and eq.\ \eqref{gen2ae}.
3-point functions of single-trace operators are one of the main 
quantities we will consider explicitly in the rest of the paper.

The scaling of 2-point functions between general multi-trace operators is more intricate because of large-$N$
factorization. For example, the 2-point function of two double-trace operators behaves at leading order as
\beq
\label{22a}
\left<k_1\,k_2\, , \barM k_3\, \barM k_4\right> \sim \left<k_1\, ,\barM k_3\right>\left<k_2\, ,\barM k_4\right>+\left<k_1\, , \barM k_4\right>\left<k_2\, ,\barM k_3\right> \sim \OO(N^0)~.
\eeq
The leading order behavior is dominated by factorization, unless the single-trace 2-point functions above vanish. 
Clearly, at this order the 2-point function \eqref{22a} does not contain any new information beyond \eqref{11a}.
However, the connected version of $\left<k_1\,k_2\, , \barM k_3\, \barM k_4\right>$,
\beq
\label{22b}
\left<k_1\,k_2\, ,\barM k_3\, \barM k_4\right>_c \equiv \left<k_1\,k_2\, ,\barM k_3\, \barM k_4\right> -\left<k_1\, ,\barM k_3\right>\left<k_2\, ,\barM k_4\right>-\left<k_1\, ,\barM k_4\right>\left<k_2\, ,\barM k_3\right> \sim \OO(N^{-2})~,
\eeq
is far more interesting and scales with a subleading power of $N$, as $\OO(N^{-2})$. The leading contribution to the
connected correlator is also determined by suitable combinations of derivatives of the free energy term $F_0$. Hence,
quantities like \eqref{22b} are also accessible within the saddle-point approximation of the matrix model and contain
useful information about the large-$N$ gauge theory. We will consider observables related to \eqref{22b} in subsection
\ref{basisindependent}.

More generally, we can consider the 2-point function of multi-trace operators
\beq
\label{eqn:genmt}
\left< k_1\cdots k_m\, ,\barM k_{m+1} \cdots \barM k_{m+n}\right>_c \equiv \left< k_1\cdots k_m\, ,\barM k_{m+1} \cdots \barM k_{m+n}\right> - \textrm{(factorized pieces)} \sim \OO(N^{2-m-n})~.
\eeq
This quantity is precisely what we would get if we started from a connected $(m+n)$-point function and took the limit
where the insertions of all the chiral operators go to infinity and the insertions of the anti-chiral operators go to
zero. Again, since these objects are expressed in terms of 2-point functions of chiral primaries, they can be computed
in terms of the matrix model free energy $F$. Their leading behavior in $1/N$ is determined by the \emph{leading} term
of the free energy, $F_0$.

\subsection{Single-trace 2- and 3-point functions}
\label{singletrace3pt}

Next let us take a closer look at the Gram-Schmidt procedure, \cite{Gerchkovitz:2016gxx}, at large $N$. It was argued in
\cite{Rodriguez-Gomez:2016ijh} that mixing between single- and multi-trace operators can be ignored for the purpose of
computing single-trace 2-point functions in flat space from the sphere correlators. As a consequence, the 2-point
functions on the plane can be easily calculated from $F_0$. Using standard formulae for the Gram-Schmidt diagonalization
in terms of matrix determinants, we thus obtain\footnote{As explained in \cite{Rodriguez-Gomez:2016ijh}, when we map
2-point functions from the sphere to the plane, we get an additional factor $4^\Delta$ coming from the conformal
mapping.}
\beq
\label{2ptflatsphere}
\langle k, \barM k \rangle
\equiv \left\langle \Tr\left[\varphi^{k}\right], \Tr\left[\barM \varphi^{k}\right]\right\rangle_{\IR^4}
= \frac{\mathrm{det}\, \mathcal{M}_k}{\mathrm{det}\, \mathcal{M}_{k-2}}~,
\eeq
where $\mathcal{M}_k$ is the $k \times k$ matrix given by
\beq
\mathcal{M}_k 
=\left\{ - \partial_{g_m} \partial_{g_n} F_0 \right\}_{m,n=k, k-2,\ldots}~.
\eeq
As a trivial check, eq.\ \eqref{2ptflatsphere} is in agreement with the examples \eqref{gen5ae}, \eqref{gen5af}.
Explicit expressions around the weakly coupled point will be presented in section \ref{weakcoupling}.

The single-trace 3-point functions can also be determined in terms of $F_0$. 
More concretely, we are interested in computing
\begin{equation}
\label{3ptfundef}
\left<k_1\, ,k_2\, ,\barM{k}_3\right> \equiv 
\left\langle \Tr\left[\varphi^{k_1}\right], \Tr\left[\varphi^{k_2}\right], \Tr\left[\barM \varphi^{k_3}\right]\right\rangle_{\IR^4}
~, ~~ k_3=k_1+k_2
~.
\end{equation}
The following (streamlined) procedure leads to the desired result. First, we perform the Gram-Schmidt orthogonalization
procedure by diagonalizing the matrix of sphere 2-point functions of single-trace operators only. This leads to the
following formal identification
\begin{equation}
\label{GSsingtrace}
O^{\mathbb{R}^4}_{k} = \sum_\ell c_k^\ell \, O^{S^4}_{\ell}~,
\end{equation}
where $c_k^k = 1$ and the remaining $c_k^\ell$'s are determined from the condition 
$\left<O^{\mathbb{R}^4}_{k_1}, \overline{O}^{\mathbb{R}^4}_{k_2} \right>_{S^4} = 0$ for $k_1 \neq k_2$. 
Our claim is that
\begin{align}
\label{3ptflatsphere}
\left<k_1\, , k_2\, , \barM{k}_3\right> & = \sum_{\ell_1,\ell_2,\ell_3} c_{k_1}^{\ell_1} \, c_{k_2}^{\ell_2} \, c_{k_3}^{\ell_3}\left<\left[O^{S^4}_{\ell_1}O^{S^4}_{\ell_2}\right], \overline{O}^{S^4}_{\ell_3}\right>_{S^4}\nonumber\\
& = \frac{1}{N}
\sum_{\ell_1,\ell_2,\ell_3} 
c_{k_1}^{\ell_1} \, c_{k_2}^{\ell_2} \, c_{k_3}^{\ell_3} \, \partial_{g_{\ell_1}}\partial_{g_{\ell_2}}\partial_{g_{\ell_3}} F_0~. 
\end{align}
The proof of this statement is presented in appendix \ref{proof}. This formula is non-trivial because in principle it
differs from the prescription presented in subsection \ref{3general}. According to that prescription , to compute
\eqref{3ptfundef} we would have to perform the Gram-Schmidt diagonalization for the operator
$\left[O^{\IR^4}_{k_1}O^{\IR^4}_{k_2}\right]$ directly, whose expression in terms of sphere operators differs from the
square of \eqref{GSsingtrace}. The reason why we can work directly with the single-trace operators at large $N$ is due
to large-$N$ factorization of correlators, as explained in appendix \ref{proof}.

\subsubsection{Mixing with multi-trace operators}
\label{mixing}

The single-trace 3-point functions are affected by the following subtlety: $R$-charge conservation implies that the only
non-vanishing 3-point functions are ``extremal'', or more specifically $k_3 = k_1+k_2$ in \eqref{3ptfundef}. This means
that the single-trace 3-point functions, unlike the 2-point functions, are sensitive to mixing with multi-trace
operators \cite{D'Hoker:1999ea}. As a consequence, different choices for the basis of operators away from the weakly
coupled point will lead to different answers for these 3-point functions.

This is best illustrated in an example. Let us consider the operator
\beq
O'_4 \equiv O_4 + \frac{\alpha(\lambda)}{N}(O_2)^2~,
\eeq
where $\alpha(\lambda)$ is an arbitrary function of the coupling constant $\lambda$. Its 2-point function at large $N$
is identical to the one for $O_4$, so it cannot be used to distinguish the two operators. However, the 3-point function
of this operator with two $O_{2}$ operators reads
\beq
\langle 2,2,4' \rangle \underset{N\gg 1}{\approx} \langle 2,2,4\rangle 
+ 2\frac{\alpha(\lambda)}{N} \langle 2,2 \rangle^2~.
\eeq
Since both terms on the r.h.s.\ contribute at the same order, $1/N$, the leading term at large $N$ for this correlator
depends on the arbitrary function $\alpha(\lambda)$. At tree level, we can explicitly check that the correlators
computed from the sphere partition function match the ones computed with Feynman diagrams in the standard trace basis
\eqref{eqn:3ptzeta3explicit}, \eqref{eqn:norm3pt}, so $\alpha(0)=0$. As we move away from the weakly coupled point,
however, it is not obvious a priori that the scheme we are employing matches the one of ordinary perturbation theory in
flat space.

There are two possibilities to get around this issue. One is to work with quantities that are manifestly free from
ambiguities related to the choice of basis. This is the approach that is described in the following subsection.
Alternatively, one can fix the basis of operators away from the weakly-coupled point using the following well-motivated
procedure. Conformal perturbation theory provides us with a preferred connection on the space of operators
\cite{Papadodimas:2009eu}. This connection can be used to \emph{parallel transport} the operators away from the
weakly-coupled point. While this procedure depends on the path chosen to connect the points on the conformal manifold,
at large $N$ a preferred path emerges, since the conformal manifold effectively becomes one-dimensional. We explain how
to implement this procedure explicitly, and provide various examples, in section \ref{transport}.

\subsection{Basis-independent 3-point functions}
\label{basisindependent}

So far we have discussed correlation functions in a specific basis of chiral operators, where both 2- and 3-point
functions are non-trivial. In general, however, it is customary to work in a different basis, where the 2-point
functions are unit normalized, and all the non-trivial information is encoded in the coupling-constant dependence of
higher-point correlators. Alternatively, we can work directly with quantities that are manifestly free from ambiguities
arising from the choice of basis. In geometric language, we can look at scalar quantities on the conformal manifold. The
simplest such quantity that can be constructed solely from the chiral ring data is
\begin{equation}
\label{eqn:csquared}
|C^{(\Delta_1,\Delta_2)}|^2 \equiv g^{\barM M_{\Delta_1} J_{\Delta_1}} C_{J_{\Delta_1}K_{\Delta_2}}^{P_{\Delta_1+\Delta_2}} g_{P_{\Delta_1+\Delta_2} \barM Q_{\Delta_1+\Delta_2}}\, C_{\barM M_{\Delta_1} \barM R_{\Delta_2}}^{*\barM Q_{\Delta+2}}   g^{\barM R_{\Delta_2} K_{\Delta_2}}~.
\end{equation}
This object is closely related to the ``properly normalized'' 3-point functions defined in \cite{Baggio:2014ioa}. 

For example, in the case of gauge group $SU(2)$, the chiral ring is generated by $O_2 = \Tr[\varphi^2]$ and the 2-point
functions in the chiral ring are given by $g_{2n} \equiv \left<O_2^n\, ,\barM O_2^n\right>$, so we have
\beq
|C^{(2m,2n)}_{SU(2)}|^2 = \frac{g_{2m+2n}}{g_{2m}\, g_{2n}} \equiv  (\hat{C}_{2m\, 2n}^{2m+2n})^2~,
\eeq
where $\hat{C}_{2m\, 2n}^{2m+2n}$ are the 3-point functions written in a basis where the 2-point functions are unity.
These quantities were computed exactly in \cite{Baggio:2014ioa}.

In order to compute \eqref{eqn:csquared} at large $N$, we would need to compute the 2-point functions of \emph{all} the
chiral operators (both single- and multi-trace) to the appropriate order in $N$ and then take the large-$N$ limit. The
leading order $\OO(N^0)$ term in $|C^{(\Delta_1,\Delta_2)}|^2$ is a combinatoric constant determined by large-$N$
factorization, so to get non-trivial results we have to consider the terms of order $1/N^2$ in \eqref{eqn:csquared}.

It is easy to see that these $1/N^2$ corrections are captured by the leading order free energy $F_0$. Indeed, we can
work in a basis where $C_{IJ}^{K} = \delta_{I+J}^K$, so that (schematically)
\begin{align}
|C^{(\Delta_1,\Delta_2)}|^2 &= g^{\barM M_{\Delta_1} J_{\Delta_1}}\, g^{\barM R_{\Delta_2} K_{\Delta_2}}\, g_{J_{\Delta_1}+K_{\Delta_2}, \barM M_{\Delta_1}+\barM R_{\Delta_2}}\\
& =  g^{\barM M_{\Delta_1} J_{\Delta_1}}\, g^{\barM R_{\Delta_2} K_{\Delta_2}}\,\left(  g_{J_{\Delta_1} \barM M_{\Delta_1}}g_{K_{\Delta_2} \barM R_{\Delta_2}} +g_{J_{\Delta_1} \barM R_{\Delta_2}}g_{K_{\Delta_2} \barM M_{\Delta_1}}+ g^c_{J_{\Delta_1}+K_{\Delta_2}, \barM M_{\Delta_1}+\barM R_{\Delta_2}}\right)~.
\end{align}
$g^c_{J_{\Delta_1}+K_{\Delta_2}, \barM M_{\Delta_1}+\barM R_{\Delta_2}}$ is the limit of the connected 4-point function
$\left<O_{J_{\Delta_1}} O_{K_{\Delta_2}} \barM O_{\barM M_{\Delta_1}}\overline{O}_{\barM R_{\Delta_2}}\right>_c$ where
the (anti-)chiral operators are sent to the same point. In the large-$N$ limit, the first two terms in the expression
above give the $\OO(N^0)$ factorized contribution to $|C^{(\Delta_1,\Delta_2)}|^2$, while the connected 4-point function
behaves as $\OO(1/N^2)$. The free energy is the generating function for the connected correlators, hence we conclude
that the $1/N^2$ correction to $|C^{(\Delta_1,\Delta_2)}|^2$ can indeed be computed (at least in principle) from the
leading term in the free energy $F_0$.

In fact, in the case where $\Delta_1 = 2$, $\Delta_2 = \Delta$, we can derive an explicit relation for the $1/N^2$
correction to $|C^{(2,\Delta)}|^2$ in terms of the single-trace 2-point functions \eqref{2ptflatsphere}. This is
possible because this quantity obeys a very simple recursive relation coming from the \tts equations that can be solved
explicitly.

\subsubsection{\tts equations and $|C^{(2,\Delta)}|^2$}

We recall the general \tts equations for a (complex) 1-dimensional moduli space
 \bea
\label{reviewaj}
&&\partial_{\barM \tau} \left( g^{\barM M_\Delta L_\Delta} \partial_\tau g_{K_\Delta \barM M_\Delta} \right)
\\
&&= C_{2K_\Delta}^{P_{\Delta+2}} g_{P_{\Delta+2} \barM Q_{\Delta+2}}\, C_{\barM 2 \barM R_\Delta}^{*\barM Q_{\Delta+2}}   g^{\barM R_\Delta L_\Delta} 
- g_{K_\Delta \barM N_\Delta} \, C_{\barM 2 \barM U_{\Delta-2}}^{*\barM N_\Delta}  g^{\barM U_{\Delta- 2}  V_{\Delta-2}} C_{2 V_{\Delta-2}}^{L_\Delta}
- g_2 \, \delta_{K_\Delta}^{L_\Delta}
~.\nonumber
\eea
If we contract the indices appropriately and define the quantity
\beq
\label{eqn:curvature}
R^{(\Delta)} \equiv (g_2)^{-1}\partial_{\barM \tau} \left( g^{\barM M_\Delta K_\Delta} \partial_\tau g_{K_\Delta \barM M_\Delta} \right)~,
\eeq
the equations \eqref{reviewaj} simply become
\beq
\label{recursion}
|C^{(2,\Delta)}|^2 = |C^{(2,\Delta-2)}|^2 + R^{(\Delta)} + n^{(\Delta)}~,
\eeq
where $n^{(\Delta)}$ is the number of chiral primary operators of dimension $\Delta$. This recursion equation can be
solved explicitly, and it gives
\beq
\label{solution}
|C^{(2,\Delta)}|^2 = \sum_{\Delta' \leq \Delta} (n^{(\Delta')} + R^{(\Delta')})~,
\eeq
where the sum runs over even conformal dimensions only.

Let us now examine how the recursion equation \eqref{recursion} and its solution \eqref{solution} behave in the
large-$N$ limit. We begin by analyzing the ``curvature'' term \eqref{eqn:curvature}. Recall that the large-$N$ limit is
taken by keeping the 't Hooft coupling constant fixed
\beq
\lambda = g^2 N = \frac{8 \pi i N}{\tau - \barM \tau}~.
\eeq 
Since instanton corrections are suppressed in this limit, we can assume that all the quantities (in particular, the
2-point functions) depend on $\lambda$ only. Therefore
\beq
R^{(\Delta)} = (g_2)^{-1}\partial_{\barM \tau} \left( g^{\barM M_\Delta K_\Delta} \partial_\tau g_{K_\Delta \barM M_\Delta} \right) = (g_2)^{-1}\frac{\lambda^2}{64 \pi^2 N^2} \partial_{\lambda} \left(\lambda^2\, g^{\barM M_\Delta K_\Delta} \partial_\lambda g_{K_\Delta \barM M_\Delta} \right) ~.
\eeq
The matrix of 2-point functions can be written as
\bea
g_{K_\Delta \barM M_\Delta} & = g_{0,K_\Delta \barM M_\Delta} + \frac{1}{N} g_{1,K_\Delta \barM M_\Delta} + \ldots,
\eea
and its inverse has a similar expansion, where the leading term is just $(g_{0,K_\Delta \barM M_\Delta})^{-1}$.
Therefore we have
\beq
R^{(\Delta)} = \frac{1}{N^2} R^{(\Delta)}_0 + \ldots~,
\eeq
where the ellipses indicate higher order terms in $1/N$ and $R^{(\Delta)}_0$ is given by
\beq
\label{eq:leadingcurvature}
R^{(\Delta)}_0 = (g_2)^{-1}\frac{\lambda^2}{64 \pi^2}  \partial_{\lambda} \left(\lambda^2\, g_{0}^{\barM M_\Delta K_\Delta } \partial_\lambda g_{0,K_\Delta \barM M_\Delta}  \right)~.
\eeq
$g_{0,K_\Delta \barM M_\Delta}$ is in turn given by the 2-point functions of single-trace operators \emph{only}, which
can be computed using \eqref{2ptflatsphere}.

We can now consider the behavior of $|C^{(2,\Delta)}|$ at large-$N$. Using \eqref{solution} we see that
\beq
\label{eqn:csquared2sol}
|C^{(2,\Delta)}|^2 = \sum_{\Delta' \leq \Delta} n^{(\Delta')} + \frac{1}{N^2} \sum_{\Delta' \leq \Delta} R^{(\Delta')}_0 + \OO(\frac{1}{N^3})~.
\eeq
It is clear that the $1/N^2$ correction to $|C^{(2,\Delta)}|^2$ can be determined directly from $F_0$ using equation
\eqref{2ptflatsphere}. We will give explicit formulae for $|C^{(2,\Delta)}|^2$ as a perturbative series in $\lambda$
around $\lambda = 0$ in section \ref{weakBasisIndependent}.

\section{Weak coupling results}
\label{weakcoupling}
In this section we analyze in detail the correlators described in the previous section around the weak coupling point
$\lambda = 0$. We begin by implementing explicitly the Gram-Schmidt diagonalization procedure at tree-level. This allows
us to compute the flat-space tree-level 2-point functions and 3-point functions of single-trace operators. At this
order, the computation is identical to $\mathcal{N}=4$ SYM and indeed we reproduce the results of \cite{Lee:1998bxa}.

We then show that the first non-trivial subleading corrections to these correlators can also be computed explicitly in
flat space, thanks to a simplifying property of the 1-loop determinant of the matrix model noticed in
\cite{Gerchkovitz:2016gxx}. Using the techniques of appendix \ref{appweak}, we also present several examples of
3-point functions computed to much higher order in $\lambda$. We also deal with the subtleties related to mixing with
multi-trace operators, anticipated in section \ref{mixing}, by introducing a notion of parallel transport of operators,
which is natural in conformal perturbation theory.

Finally, we analyze the squared structure constants $|C^{(2,\Delta)}|^2$ defined in \eqref{eqn:csquared}; we explicitly
compute the leading and subleading (order $\lambda^2$) results for general $\Delta$ and we present results to higher
orders in $\lambda$ in examples.

\subsection{Single trace 2- and 3-point functions at tree-level}
\label{treelevel}

Here we consider the Gram-Schmidt procedure at large $N$ and at tree-level in $\lambda$. 
For simplicity we present the
procedure for the sector of even chiral primaries. We start with the matrix of 2-point functions of single-trace
operators on the sphere, which is given by \cite{Rodriguez-Gomez:2016ijh}
\beq
\left< O_{2k_1}^{S^4}\, ,\barM{O}^{S^4}_{2k_2}\right>_{S^4} = \left({\lambda \over 4\pi}\right)^{k_1+k_2}{ \Gamma\left(k_1+{1\over 2}\right) \Gamma\left(k_1+{1\over 2}\right) \over \pi (k_1+k_2)\Gamma(k_1) \Gamma(k_2) }~.
\eeq
We will also need the 3-point functions on the sphere \cite{Rodriguez-Gomez:2016ijh}
\beq
\left< O_{2k_1}^{S^4}\, , O_{2k_2}^{S^4}\, ,\barM{O}_{2k_3}^{S^4}\right>_{S^4}   = \left({\lambda \over 4\pi} \right)^{k_1+k_2+k_3}{ \Gamma\left(k_1+{1\over 2}\right)  \Gamma\left(k_2+{1\over 2}\right) \Gamma\left(k_3+{1\over 2}\right)
\over \pi^{3/2}\Gamma(k_1) \Gamma(k_2)\Gamma(k_3)}~.
\eeq
A more general formula for even and odd chiral primary three-point functions on the sphere is given in appendix
\ref{appweak}. Performing the Gram-Schmidt procedure we find
\beq
O_{2k}^{{\IR}^4} = \sum_\ell c_k^\ell O_{2\ell}^{S^4}
\eeq
with
\beq
c_k^\ell = -2k \left(-{\lambda \over 16\pi}\right)^{k-\ell}  {\Gamma(k+\ell) \over \Gamma(2\ell+1)\Gamma(k-\ell+1)}~.
\eeq
These coefficients agree with the Chebyshev prescription of \cite{Rodriguez-Gomez:2016cem}. We can then use our previous
formulae to compute 2- and 3-point functions on ${\mathbb R}^4$. We find
\beq
\left<2k\, , 2\barM k\right> = 2 k \left( {\lambda \over 16 \pi}\right)^{2k}~,
\eeq
\beq
\left<2k_1\, , 2k_2\, , 2 \barM k_1+2 \barM k_2\right> = 8 k_1 k_2  (k_1+k_2) \left({\lambda \over 16 \pi}\right)^{2 k_1+ 2k_2} 
\eeq
If we normalize canonically the 2-point functions, we find
\beq
\left<2k_1\, , 2k_2\, , 2 \barM k_1+2 \barM k_2\right> = 2\sqrt{2k_1 k_2 (k_1+k_2)}~,
\eeq
which agrees with the results of \cite{Lee:1998bxa}.

\subsection{Single-trace 2- and 3-point functions at higher orders in $\lambda$}
\label{23higher}

Using \eqref{2ptflatsphere} and \eqref{3ptflatsphere}, it is straightforward to explicitly compute the 2- and 3-point
functions up to arbitrarily high order in $\lambda$. Remarkably, it is possible to obtain a closed form expression for
the 2- and 3-point functions in flat space to second order in $\lambda$. Indeed, it was noticed in
\cite{Gerchkovitz:2016gxx} that the 1-loop determinant of the matrix model expanded around $a=0$ (for finite $N$) is
given by
\begin{equation}
\label{eqn:GKzeta3prop}
\ZZ_{1-loop}(a) = 1 - 3 \zeta(3) (\Tr a^2)^2 + \ldots~,
\end{equation}
which means that the $\zeta(3)$ correction to a correlation function can be obtained from the tree-level result by
applying an appropriate derivative operator. We now find this operator explicitly and study its behavior at large $N$.

We can rewrite \eqref{eqn:GKzeta3prop} in the form
\begin{equation}
\llangle A\rrangle \approx \frac{1}{\ZZ}\left(1 -  3 \zeta(3) \frac{\partial^2}{\partial(2\pi \mathrm{Im} \tau)^2}\right) \ZZ_0[A]~,
\end{equation}
where $\ZZ_0[A]$ is the matrix model partition function at tree-level with the insertion of the operator $A$. Notice
that in converting the expansion of the 1-loop determinant into a derivative, we have assumed that $A$ itself does not
depend on $\tau$ in its representation inside the integral. In particular, this means that this equation cannot be
applied directly to connected correlators, since the mixing of an operator with the identity on the sphere will
exhibit a non-trivial dependence on $\tau$. It is easy to show that the equation above is equivalent to
\begin{equation}
\llangle A\rrangle \approx \llangle A \rrangle_0 -  3 \zeta(3) \left(\frac{2}{\ZZ_0}\bd \ZZ_0\, \bd \llangle A \rrangle_0 +\bd^2 \llangle A\rrangle_0\right)~,
\end{equation}
where $\llangle A\rrangle_0$ is the expectation value of $A$ at tree-level and we have defined
\begin{equation}
\bd \equiv \frac{\partial}{\partial(2\pi \mathrm{Im} \tau)}~
\end{equation}
to avoid clutter in the equations. A connected correlator will then satisfy
\begin{align}
\llangle A_1 A_2\rrangle_c  & \equiv \llangle A_1 A_2\rrangle - \llangle A_1\rrangle \llangle A_2\rrangle \nonumber\\
&  \approx \llangle A_1 A_2\rrangle_{0,c} -  3 \zeta(3) \left(\frac{2}{\ZZ_0}\bd \ZZ_0\, \bd  \llangle A_1 A_2\rrangle_{0,c} + \bd^2\llangle A_1 A_2\rrangle_{0,c} + 2\bd \llangle A_1\rrangle_0\, \bd \llangle A_2\rrangle_0\right)~.
\end{align}
In the large-$N$ limit, only the first and third terms in the brackets will contribute, since they scale like $N^0$
while the second term scales like $N^{-2}$.

Translating these formulae to the language of connected correlation functions on the sphere we have
\bea
\label{eqn:2ptzeta3}
\left< O_{k_1}^{S^4}\, ,\barM{O}^{S^4}_{k_2}\right>_{S^4} &\approx& 
\left< O_{k_1}^{S^4}\, ,\barM{O}^{S^4}_{k_2}\right>_{S^4,0}  
\cr
&&-  \frac{6}{\pi^2} \zeta(3) \left(\frac{\lambda^3}{128 \pi^2}\partial_{\lambda}  \left< O_{k_1}^{S^4}\, ,\barM{O}_{k_2}^{S^4}\right>_{S^4,0} + \left<O_2^{S^4}\, , {O}_{k_1}^{S^4}\right>_{S^4,0}\,  \left<O_2^{S^4}\, , \barM{O}_{k_2}^{S^4}\right>_{S^4,0}\right)~.\nonumber\\
\eea
Analogously, one can derive the $\zeta(3)$ correction to the connected 3-point function as
\bea
\label{eqn:3ptzeta3}
&&\left< O_{k_1}^{S^4}\, , O_{k_2}^{S^4}\, ,\barM{O}_{k_3}^{S^4}\right>_{S^4} 
\approx \left< O_{k_1}^{S^4}\, ,O_{k_2}^{S^4}\, ,\barM{O}_{k_3}^{S^4}\right>_{S^4,0} 
\cr
&&- \frac{6}{\pi^2} \zeta(3) \left(\frac{\lambda^3}{128 \pi^2}\partial_{\lambda}  
\left< O_{k_1}^{S^4} \, , O_{k_2}^{S^4} \, ,\barM{O}_{k_3}^{S^4}\right>_{S^4,0} 
+ \left<O_2^{S^4}\, , \barM{O}_{k_1}^{S^4}\right>_{S^4,0}\, \left<O_2^{S^4} \, , O_{k_2}^{S^4} \, ,\barM{O}_{k_3}^{S^4}\right>_{S^4,0} + \mathrm{permutations}\right)~.\nonumber\\
\eea

Remarkably, these formulae are valid even for flat-space correlators. To prove this, consider for definiteness the case
of 2-point functions \eqref{eqn:2ptzeta3} applied to the case $k_1=k_2=k$. Employing the Gram-Schmidt procedure as in
section \ref{fromStoR}
\begin{align}
O_{k}^{\IR^4}  = \sum_{i \leq k} c_i O_{i}^{S^4}~,~~
\barM{O}_{k}^{\IR^4}  = \sum_{j\leq k} \bar c_j \barM O_{j}^{S_4}
\end{align}
we deduce that the leading perturbative correction to the flat-space 2-point function can be written as
\bea
\delta \left<{k} \, ,\overline{k}\right>_{\IR^4} 
&=& \sum_{i,j} c_i \bar c_j \, \delta\left<O_i^{S^4}\, , \barM{O}_j^{S^4}\right>_{S^4} 
+ \sum_i \delta c_i \left< O_i^{S^4}\, ,\overline{{O}}_{k}^{\IR^4}\right>_{S^4} 
+ \sum_j \delta \bar c_j \left<{O}_{k}^{\IR^4}\, , \barM{O}_{j}^{S^4}\right>_{S^4} 
\cr
&=& \sum_{i,j} c_i \bar c_j \delta\left< O_i^{S^4} \, ,\barM{O}_{j}^{S^4}\right>_{S^4}~,
\eea
where in the last equality we used that the operators $O_k^{\IR^4}$ are orthogonal to all the operators of lower
conformal dimension by definition. Applying \eqref{eqn:2ptzeta3} to the previous expression, and using the same argument
to move the derivative in front of the sum, we find
\begin{equation}
\left<k\, ,\overline{k}\right> \approx\left<k\, ,\overline{k}\right>_0  
-  \frac{6}{\pi^2} \zeta(3) \left(\frac{\lambda^3}{128 \pi^2}\partial_{\lambda} \left<k\, ,\overline{k}\right>_0 
+ \left<2\, ,\overline{k}\right>_0\,  \left<2\, , \overline{k}\right>_0\right)~.
\end{equation}

The same argument applies to the 3-point functions as well. Consequently, if we know the tree-level 2- and 3-point
functions on the plane, we can determine their $\zeta(3)$ correction using \eqref{eqn:2ptzeta3} and
\eqref{eqn:3ptzeta3}. The final result is
\begin{align}
\left<k \, , \barM{k} \right> & = k \left(\frac{\lambda}{16 \pi}\right)^k \left(1 
-\frac{3\, \zeta(3)}{4 (2\pi)^4}\left( k + \delta_{k,2}\right) \lambda^2 
+ \ldots \right)~,\\
\label{eqn:3ptzeta3explicit}
\left<k_1\, , k_2\, ,\barM{k}_3\right> & = k_1\,k_2\,k_3 \left(\frac{\lambda}{16\pi}\right)^{(k_1+k_2+k_3)/2} \times
\\
&~~~
\left( 1- \frac{3\, \zeta(3)}{4 (2\pi)^4} \left( \frac{k_1+k_2+k_3}{2} + \delta_{k_1,2}+\delta_{k_2,2}+\delta_{k_3,2} \right)
\lambda^2
+ \ldots \right)
~, \nonumber
\end{align}
\bea
\label{eqn:norm3pt}
\left<k_1\, , k_2\, , \barM{k}_3\right>_{n} &\equiv& 
\frac{\left<k_1\, ,k_2\, ,\barM{k}_3\right>}{\sqrt{\left<k_1 \, ,\barM{k}_1\right>\left<k_2\, , \barM{k}_2\right>\left<k_3\, , \barM{k}_3\right>}} 
\cr
&=&\sqrt{k_1\,k_2\,k_3} \left(1 - (\delta_{k_1,2}+\delta_{k_2,2}+\delta_{k_3,2})\frac{3 \zeta(3)}{8 (2\pi)^4} \lambda^2 + \ldots \right)~.
\eea

We first notice that the tree-level results match the ones computed in \cite{Lee:1998bxa}, as they should. Furthermore,
the $\lambda^2$ corrections only appear when one of the operators is $O_2=\Tr[\varphi^2]$. This property does not hold
at higher orders in $\lambda$. Indeed, it is easy to use \eqref{3ptflatsphere} to explicitly compute some examples of
3-point functions to higher order in $\lambda$, even though we do not have a closed form expression for them valid for
arbitrary operators:
\begin{align}
\left<2\, , 2\, ,\barM{4}\right>_n & = 4 \left(1 - \frac{3\zeta(3)}{64\pi^4}\lambda^2 + \frac{45 \zeta(5)}{512 \pi^6}\lambda^3 + \frac{3(72 \zeta(3)^2 - 1085\zeta(7))}{32768\pi^8}\lambda^4 + \ldots\right)~,\\
\left<2\, ,4\, ,\barM{6}\right>_n & =4\sqrt{3}\left(1 - \frac{3\zeta(3)}{128\pi^4}\lambda^2 + \frac{15 \zeta(5)}{256 \pi^6}\lambda^3 + \frac{99 \zeta(3)^2 - 2275\zeta(7)}{32768\pi^8}\lambda^4 + \ldots\right)~,\\
\left<4\, ,4\, ,\barM{8}\right>_n & =8\sqrt{2} \left(1 + \frac{15 \zeta(5)}{512 \pi^6}\lambda^3 - \frac{665\zeta(7)}{16384\pi^8}\lambda^4 + \ldots\right)~,\\
\left<4\, , 6\, ,\barM{10}\right>_n & =4\sqrt{15} \left(1 + \frac{15 \zeta(5)}{512 \pi^6}\lambda^3 - \frac{35(263520 \zeta(3)^2 - 501551\zeta(7))}{32768\pi^8}\lambda^4 + \ldots\right)~,\\
\ldots & \nonumber
\end{align}
In particular, notice that while the $\lambda^2$ corrections are absent in the normalized 3-point functions that do not
involve the operator $O_2$, the $\lambda^3$ and higher corrections are present.

\subsection{3-point functions in the parallel transported basis}
\label{transport}

As we anticipated in section \ref{mixing}, the 3-point functions defined above are sensitive to mixing with multi-trace
operators. At the weakly coupled point, the canonical trace basis is particularly convenient, and we can ask if this
basis can be extended (or transported) in a canonical way across the conformal manifold. Since conformal perturbation
theory defines a natural connection on the space of operators, it is sensible to use this connection to parallel
transport the canonical trace basis defined at the weakly coupled point to other points in the conformal manifold. In
general, this is an ambiguous operation, because the parallel transported basis will depend on the particular path
chosen due to curvature. Fortunately, at large $N$, the conformal manifold becomes effectively 1-dimensional, and a
preferred path emerges, the one ``along $\lambda$''.

Let us define the vielbein-like objects $e_k^I(\lambda)$ such that
\beq
e_k^I(\lambda) O_I \propto O_k + \OO(\lambda^2)~.
\eeq
Here the index $I$ is allowed to run over all the chiral primaries (both single- and multi-trace) of dimension $\Delta =
k$ and the arbitrary proportionality constant can be chosen so that the diagonal part of the 2-point functions is unity.
A choice of $e_k^I(\lambda)$ corresponds to a choice of basis of chiral primaries that agree (up to an overall
normalization) with the tree-level trace basis when $\lambda = 0$. The parallel transported basis will be determined by
demanding that
\beq
\label{paralleltransport}
\nabla_\lambda e_k^I(\lambda) = 0~,
\eeq
where $\nabla_\lambda$ is the covariant derivative along $\lambda$. The parallel transported 3-point functions will then
be given by
\beq
\label{eqn:partran3pt}
e_{k_1}^I(\lambda)\,e_{k_2}^J(\lambda)\,e_{k_3}^K(\lambda) \left<O_I \, ,O_J\, ,\barM O_K \right>~.
\eeq
It is clear that these correlators will agree with \eqref{eqn:norm3pt} at leading order in $\lambda$.

In order to implement this procedure explicitly in the present case, we need to determine the connection from the $S^4$
partition function. To do so, we will \emph{assume} that the basis of operators on the plane implicitly defined by the
Gram-Schmidt procedure is a \emph{holomorphic basis}.\footnote{Strictly speaking, the basis that we are considering is
holomorphic only when the operators are multiplied by the factor $e^{\frac{R}{c}\mathcal{K}}$, where $\mathcal{K}$ is
the K\"ahler potential of the theory. For more details and a discussion on the effect of K\"ahler ambiguities, we refer
the reader to \cite{Baggio:2014ioa}.} This assumption passes an important consistency check, namely that the resulting
2-point functions do obey the \tts equations written in a holomorphic basis \cite{Gerchkovitz:2016gxx}. However, as we
noted previously at the end of section \ref{fromStoR}, we are not aware of a complete proof of this statement. If the
assumption is correct, then it is easy to show \cite{Baggio:2014ioa} that the connection along $\lambda$ is
\beq
\nabla_\lambda = \partial_\lambda + \frac{1}{2} g^{-1} \partial_\lambda g~,
\eeq
where $g$ is the matrix of 2-point functions at the appropriate level. Using this and the explicit results up to order
$\lambda^2$ of the previous subsection, it would be easy to compute \eqref{eqn:partran3pt} up to this order. However,
due to the property in equation \eqref{eqn:GKzeta3prop}, this correction will vanish.

To exemplify how the procedure works in practice, we explicitly work out the case $k_1 = k_2 = 2$, $k_3 = 4$ up to order
$\lambda^3$. The relevant matrices of 2-point functions are then
\begin{align}
g^{(2)} & = 
2\left(\frac{\lambda}{16 \pi}\right)^2 \left(1 - \frac{9\, \zeta(3)}{64 \pi^4} \lambda^2 +  \frac{15\, \zeta(5)}{128 \pi^6} \lambda^2  + \ldots \right)~,\\
g^{(4)} & =4\left(\frac{\lambda}{16 \pi}\right)^4  \begin{pmatrix*}[l]
2 - \frac{9 \zeta(3)}{16 \pi^4}\lambda^2  + \frac{15 \zeta(5)}{32 \pi^6}\lambda^3 + \ldots  &
\frac{1}{N}\left(4 - \frac{9 \zeta(3)}{8 \pi^4}\lambda^2  + \frac{145 \zeta(5)}{128 \pi^6}\lambda^3 + \ldots \right) \\[5px]
\frac{1}{N}\left(4 - \frac{9 \zeta(3)}{8 \pi^4}\lambda^2  + \frac{145 \zeta(5)}{128 \pi^6}\lambda^3 + \ldots \right) &
1 - \frac{3 \zeta(3)}{16 \pi^4}\lambda^2  + \frac{5 \zeta(5)}{32 \pi^6}\lambda^3 + \ldots 
\end{pmatrix*}~.
\end{align}
We now have everything we need to solve \eqref{paralleltransport}. The result is
\begin{align}
\frac{\sqrt{2} \lambda}{16\pi} e_2 (\lambda) & =  1 + \frac{9 \zeta(3)}{128 \pi^4}\lambda^2 - \frac{15 \zeta(5)}{256 \pi^6} \lambda^3 + \ldots~,\\
\frac{2\lambda^2}{(16\pi)^2} e^I_4 (\lambda) & = \begin{pmatrix} \frac{3 \zeta(3)}{32 \pi^4\,N}\lambda^2 - \frac{65 \zeta(5)}{512 \pi^6\,N} \lambda^3 \\[5px] 1 +  \frac{3 \zeta(3)}{32\pi^4}\lambda^2 - \frac{5 \zeta(5)}{64 \pi^6} \lambda^3 \end{pmatrix} + \ldots ~.
\end{align}
We notice that $e_{4}$ acquires a component along the multi-trace chiral primary $(O_2)^2$ at order $\lambda^2$. This
means that as we parallel transport the operator $O_4$ to non-zero $\lambda$, it mixes with $(O_2)^2$. This mixing is a
$1/N$ effect, as expected.

Finally, the 3-point function in the parallel transported basis reads
\beq
e_2\, e_2\, e^I_4 \left<O_2 \, , O_2\, ,\barM O_I\right> = \frac{1}{N} \left(4 + \frac{25 \zeta(5)}{256 \pi^6} \lambda^3 + \OO(\lambda^4)\right)~.
\eeq
The tree-level piece is of course the same as before. We also notice that the $\lambda^2$ correction is absent, as
expected from \eqref{eqn:GKzeta3prop}, but the correlator does receive quantum corrections from order $\lambda^3$ and
higher.

\subsection{Basis-independent 3-point functions}
\label{weakBasisIndependent}

In this final subsection, we study the objects $|C^{(2,\Delta)}|^2$ defined in \eqref{eqn:csquared} at weak coupling. We
start by giving an analytic expression of the curvatures $R_0^{(\Delta)}$, defined in \eqref{eq:leadingcurvature}, to
order $\lambda^2$. Since only the leading order $\OO(N^0)$ part of the metric contributes, the problem simplifies
considerably. First of all, the metric is diagonal. Also, the diagonal elements are either single-trace 2-point
functions or multi-trace 2-point functions that factorize into the product of single-trace 2-point functions at leading
order in $1/N$. The curvature for the single-trace part is given by
\begin{equation}
R^{k}_0 \equiv \frac{1}{\left<{2}\, ,\bar{2}\right>} 
\partial_\tau \partial_{\bar\tau} \log \left<k \, , \bar{k}\right> = 2\, k - \frac{9\, k\, \zeta(3)}{32 \pi^4} \lambda^2 - \delta_{k,2}\frac{9\,\zeta(3)}{16 \pi^4} \lambda^2 + \ldots~.
\end{equation}
The piece of the curvature coming from the 2-point functions of the operators $\prod_k k^{n_k}$ will then be given by
\begin{equation}
R^{\{n_k\}}_0 = \sum_k n_k R^{k}_0~,
\end{equation}
where we used large-$N$ factorization. Therefore, the leading large-$N$ contribution to the curvature on the space of
chiral primaries of conformal dimension $\Delta$ is given by
\begin{equation}
R^{(\Delta)}_0 = \sum_{\mathclap{\substack{\{n_k\} \\ \sum_{k} n_k  k = \Delta}}} \sum_k  n_k R^{k}_0 =  2\,\Delta\, n_{(\Delta)}  - \left( \Delta\, n_{(\Delta)}+ 2\, p_{(\Delta)}\right)\frac{9\, \zeta(3)}{32 \pi^4} \lambda^2 + \ldots~,
\end{equation}
where $n_{(\Delta)}$ is the dimension of the space of chiral primaries at level $\Delta$
(\url{https://oeis.org/A182746}) and $p_{(\Delta)}$ is the number of 2's in all partitions of $\Delta$ that do not
contain 1 as part (\url{https://oeis.org/A182716}). Combining this result with \eqref{eqn:csquared2sol} gives us the
$1/N^2$ piece to order $\lambda^2$ exactly. As before, it is very easy to compute higher-order corrections by following
the algorithm of section \ref{singletrace3pt}, but we do not have a closed form expression for them.

The case where $\Delta = 2$ is particularly interesting, as it gives the leading contribution to the conformal block
expansion of the 4-point function
\beq
\left<\phi_2(x_1)\, \phi_2(x_2)\, \barM \phi_2(x_3)\,\barM \phi_2(x_4)\right>
\eeq
in the chiral channel, namely $x_1 \to x_2$, $x_3 \to x_4$. In the case of $\mathcal{N}=4$ SYM, the analogous object was
studied both from the field theory side \cite{Bianchi:1999ge} and the gravity side \cite{Arutyunov:2000ku}, and the
results turned out to be independent of the coupling constant consistent with the $\NN = 4$ non-renormalization theorem
\cite{Lee:1998bxa,D'Hoker:1998tz,D'Hoker:1999ea,Intriligator:1998ig,Intriligator:1999ff,Eden:1999gh,Petkou:1999fv,Howe:1999hz,Heslop:2001gp,Baggio:2012rr}.
In our case, we can explicitly see that this quantity does depend non trivially on the coupling constant, its first
perturbative corrections around $\lambda=0$ being
\beq
|C^{(2,2)}|^2 = 2 + \frac{1}{N^2} \left(4 - \frac{9 \zeta(3)}{8\pi^4}\lambda^2 + \frac{75\zeta(5)}{32\pi^6}\lambda^3 + \frac{9(27\zeta(3)^2 - 350\zeta(7))}{1024\pi^8}\lambda^4 + \OO(\lambda^5)\right) + \OO(N^{-3})~.
\eeq
This result is consistent with the lower bound
\beq
\label{boot}
|C^{(2,2)}|^2 \geq 2 + \frac{2}{3c}
\eeq
in \cite{Beem:2014zpa} that follows from conformal bootstrap techniques.

\section{Partial strong coupling results}
\label{strong}

The analysis of 2- and 3-point functions at strong coupling requires a solution of the large-$N$ saddle-point equation
of the matrix model at large $\lambda$. The saddle point equation has the qualitative form
\be
\label{integraleqb}
\int_{\mu_-}^{\mu_+} dx  \left[{1\over x-y} - K(x-y)\right] \rho(y) = {8 \pi^2 \over \lambda}x - K(x) + \sum_{n=2}^M t_n x^n
~.\ee 
Details about the function $K(x)$, as well as the meaning of the couplings $t_n$ can be found in appendix 
\ref{deformedmatrix}. We are interested in a single-cut solution with the density of eigenvalues $\rho(x)$ 
supported in the interval $[\mu_-,\mu_+]$. Unfortunately, we have not been able to find an analytic solution 
of the integral equation \eqref{integraleqb} at arbitrary finite values of the
couplings. However, in \cite{Passerini:2011fe} it was argued that $\mu = \mu_+=-\mu_- \to \infty$ in the strong 
coupling limit, $\lambda \rightarrow \infty$, at $t_n=0$. It was further argued that the leading order relation 
between $\lambda$ and $\mu$ in this limit is
\beq
\label{strongab}
\mu \simeq \frac{2}{\pi} \log \lambda~, ~~ \lambda \gg 1
~.\eeq
The presence of small higher single-trace couplings $t_n$ $(n\geq 2)$ will not affect this qualitative behavior of 
$\mu_\pm$, but the specifics of the dependence of $\mu_\pm$ on the general $t_n$, that generalizes 
\eqref{strongab}, requires a careful analysis of the integral equation \eqref{integraleqb}.

Ref. \cite{Passerini:2011fe} further proposed an approximate analysis of \eqref{integraleqb} based on the Wiener-Hopf
method. Details of this approach, suitably generalized to include the effects of the couplings $t_n$, are presented in
appendix \ref{appstrong}. By running the approximate Wiener-Hopf method for the saddle-point equations, evaluating
matrix model correlation functions and performing the eventual Gram-Schmidt orthogonalization procedure we can obtain
approximate results for 2- and 3-point functions of single-trace operators in the $\IR^4$ theory. For example, in this
way we obtain the following large-$\lambda$ behavior of the correlation functions \eqref{gen5ae},\footnote{The behavior
of $\left\langle \Tr\left[\varphi^2 \right], \Tr\left[\barM\varphi^2 \right] \right\rangle_{\IR^4}$ at strong coupling,
determined by the same approximation method, was reported also in \cite{Rodriguez-Gomez:2016ijh} and agrees with our
results below.} \eqref{gen5af},
\eqref{gen5ad}
\beq
\label{strongac}
\left\langle \Tr\left[\varphi^2 \right],  \Tr\left[\barM\varphi^2 \right] \right\rangle_{\IR^4}
\sim \left( \log \lambda \right)^2
~,\eeq
\beq
\label{strongad}
\left\langle \Tr\left[\varphi^4 \right],  \Tr\left[\barM\varphi^4 \right] \right\rangle_{\IR^4} 
\sim \left( \log \lambda \right)^6
~,\eeq
\beq
\label{strongae}
\left\langle \Tr\left[\varphi^{2}\right], \Tr\left[\varphi^{2}\right], \Tr\left[\barM \varphi^{4}\right]\right\rangle_{\IR^4}
\sim \lambda \left( \log \lambda \right)^3
~.\eeq
Many more explicit results like this can be obtained from the computations of appendix \ref{appstrong}.

Determining the precise numerical prefactor in these expressions is hard. 
As we detail in appendix \ref{appstrong} the employed Wiener-Hopf approximations are not
based on a well-controlled expansion in terms of a parametrically small number. In fact, performing this computation 
at the next iteration we found corrections to the leading order coefficients that are numerically comparable to the 
leading contribution.

In search of an independent check of the leading large-$\lambda$ scaling of correlation functions obtained with
the Wiener-Hopf method, let us consider the connected 2-point functions 
of single-trace operators in the matrix model. Using the independent results of appendix \ref{2density}, in particular
equations \eqref{conDensaa} and \eqref{conDensdc}, we focus on the following contribution to the general 
connected 2-point function in the matrix model
\beq
\label{strongaf}
\langle \Tr [\varphi^n] \Tr[\varphi^m] \rangle_c \sim \int_{-\mu}^\mu dx \int_{-\mu}^\mu dy \, \barM \rho_2 (x,y)
x^{n} y^{m}  +\ldots
~.\eeq
The dots indicate corrections from the integration of subleading terms in the density of connected 2-point functions 
$\bar\rho_2(x,y)$. We assume
that such contributions either exhibit the same scaling in the large-$\mu$ limit or subleading scaling.
In the case of the standard matrix model, where the exact two-point function density is known (see eq.\
\eqref{conDensca}) we can check that these corrections exhibit the same large-$\mu$ scaling as \eqref{strongaf}
correcting the numerical coefficients one finds from \eqref{strongaf}. 

In any case, focusing on the leading scaling of the term on the r.h.s.\ of \eqref{strongaf} we find
\bea
\label{strongag}
&&\int_{-\mu}^\mu dx \int_{-\mu}^\mu dy \, \barM \rho_2 (x,y) x^{n} y^{m}
= \mu^{n+m+2} \int_{-1}^1 dx \int_{-1}^1 dy \, \barM \rho_2 (\mu x, \mu y) x^{n} y^{m}
\cr
&&\sim - \frac{3}{\sqrt{2} \pi^{3/2}} \mu^{n+m+2}  \int_{-1}^1 dx \int_{-1}^1 dy \,  \frac{x^{n} y^{m}}{(\mu x-\mu y)^4}
\propto \mu^{n+m-2}
~.
\eea
To obtain the second line we used the large-$\mu$ asymptotics of the expression \eqref{conDensdc}
at finite non-vanishing $x-y$. 
If \eqref{strongag} is indeed a term that contributes to the leading large-$\mu$ scaling of 
$\langle \Tr [\varphi^n] \Tr[\varphi^m] \rangle_c$ we deduce that
\beq
\label{strongai}
\langle \Tr [\varphi^n] \Tr[\varphi^m] \rangle_c \sim \OO( \mu^{n+m-2} )
=\OO( \left(\log \lambda\right)^{n+m-2} )
~.\eeq
This prediction agrees well with the large-$\mu$ scaling obtained with the approximate Wiener-Hopf method.
This is partially reassuring. 

Let us also note in passing that the result \eqref{strongac} is trivially consistent with the bootstrap bound
\eqref{boot}. The coefficient of the $1/N^2$ correction to $|C^{(2,2)}|^2$ is a large positive number scaling as
$\lambda^2/(\log \lambda)^3$. The $\lambda^2$ factor is consistent with the linear $\lambda$ factor in
\eqref{strongae}.

It would be very interesting to obtain a better handle on the precise numerical coefficients in front of the above 
scalings of the 2- and 3-point functions,
either by improving the Wiener-Hopf method, or by developing further the computation of connected 2- and
higher-$n$-point function densities (we refer the reader to appendix \ref{2density} for additional comments
on this approach).

\subsection*{Acknowledgments}
We would like to thank N. Bobev, A. Bzowski, P. Heslop, Z. Komargodski, N. Mekareeya, S. Minwalla, E. Pomoni, D.
Rodriguez-Gomez, J. Russo, M. Schillo, A. Stergiou, G. Tartaglino-Mazzucchelli, B. van Rees for useful discussions. The
work of MB is supported in part by the European Research Council grant no. ERC-2013-CoG 616732 HoloQosmos and in part by
the Belgian National Science Foundation (FWO) and the European Union's Horizon 2020 research and innovation programme
under the Marie Sk\l odowska-Curie grant agreement no.\ 665501. MB is an FWO [PEGASUS]${}^2$ Marie Sk\l odowska-Curie
Fellow. The work of VN was partially supported by the Advanced ERC grant SM-grav 669288. KP and GV would like to thank the
Royal Netherlands Academy of Sciences (KNAW).

\begin{appendix}

\section{Deformed matrix integrals from supersymmetric localization}
\label{deformedmatrix}

In section \ref{S4correspondence} we reviewed how extremal correlation functions are related to derivatives of 
the free energy of the $\NN=2$ theory on $S^4$. Via supersymmetric localization this is also the free energy of a 
corresponding matrix model. In the case of the $SU(N)$ $\NN=2$ SCQCD theory 
the deformed partition function of interest is (after localization) \cite{Pestun:2007rz}
\beq\label{freeba}
\ZZ_{S^4} = \int d^N a \, \delta \left(\sum_{i=1}^N a_i\right) \prod_{i<j} 
\bigg[ (a_i-a_j)^2 H^2 (a_i - a_j) \bigg]
\bigg | 
e^{i \sum_{n=2}^N \pi^{n/2} \tau_n \sum_{i=1}^N (a_i)^n} 
\bigg |^2
e^{-2 N \sum_{i=1}^N \log H(a_i)}
 | \ZZ_{inst} |^2
 ~.\eeq
 The function $H(x)$ is defined as
\beq\label{freebab}
H(x) = \prod_{n=1}^\infty \left( 1+ \frac{x^2}{n^2} \right)^n e^{-\frac{x^2}{n}}
~.\eeq
$\ZZ_{inst}$ are instanton contributions.
The integral is performed over the $N-1$ elements of the $SU(N)$ Cartan subalgebra. In the $U(N)$ theory there
is no $\delta$-function restriction on the $N$ elements of the Cartan subalgebra. From now on, and 
in appendices \ref{appweak}, \ref{appstrong}, it is more convenient to work in the $U(N)$ matrix model. Eventually, 
the translation of the matrix model results to the SCFT on $\IR^4$ are identical in the $SU(N)$ and $U(N)$ cases
as far as the leading large-$N$ contribution to correlators is concerned.

\subsection{Large-$N$ limit and the saddle-point equations}

In the large-$N$ limit we can further set the instanton contributions $\ZZ_{inst}\to 1$, 
and express the single-trace couplings 
$\tau_n$ in terms of their 't Hooft combinations \eqref{scaledcouplings}. Since the end result depends only on 
the imaginary part of the couplings $\tau_n$ we set 
\beq\label{freebb}
g_n \equiv \frac{2}{N} {\rm Im} \tau_n
~.\eeq
Then, the large-$N$ $(U(N))$ partition function takes the form
\bea\label{freebc}
\ZZ_{S^4} =  e^{-N^2 \FF(\{g_n\})} = 
\int d^N a 
\prod_{i< j} 
\bigg[ (a_i-a_j)^2 H^2 (a_i - a_j) \bigg]
e^{-N \sum_{i=1}^N  \left[ \sum_{n=2}^N g_n \, \pi^{n/2} (a_i)^n + 2 \log H(a_i) \right] }
.\eea

In the large-$N$ limit it is also convenient to introduce the density of eigenvalues
\beq\label{freebd}
\rho(x) = \frac{1}{N} \sum_i \delta(x-a_i)
~,
\eeq
which is normalized so that 
\beq\label{freebe}
\int dx\, \rho(x) = 1
~.
\eeq
Below we will deal exclusively with single-cut solutions, 
where the saddle-point eigenvalues are located in a single connected 
interval $[\mu_-,\mu_+]$.

The saddle-point equations of the integral \eqref{freebc} are a special case of the general integral equation
\beq\label{freebf}
\dashint dy \, \rho(y) \left( \frac{1}{x-y} - K(x-y) \right) = f(x)
~,\eeq
where
\beq\label{freebg}
K(x) = - \frac{H'(x)}{H(x)} = 2x \sum_{n=1}^\infty \left( \frac{1}{n} - \frac{n}{n^2+x^2} \right)
~,\eeq
\beq\label{freebi}
f(x) = \frac{1}{2} \frac{dV}{dx} - K(x)~, ~~
V(x) = \sum_{n=2}^\infty \pi^{n/2}g_n x^n
~.
\eeq 
$\dashint$ denotes a principal value integral. The function $K$ controls the measure of the matrix integral, and the function $f$ controls its 
potential.

The undeformed version of these equations, with $g_n=0$ for all $n>2$, was analyzed previously in 
\cite{Passerini:2011fe}. An analysis of the deformed equations, with $g_n=0$ for all odd $n$, was initiated 
more recently in \cite{Rodriguez-Gomez:2016ijh}.
We revisit this analysis below and extend it in several directions.

Following \cite{Passerini:2011fe}, it is possible, for a general single-cut configuration, to recast the saddle-point equations 
\eqref{freebf} into the form
\bea\label{freebj}
\rho(x) &=& -  \frac{1}{\pi^2} \dashint_{\mu_-}^{\mu_+} \frac{dy}{x-y} \sqrt{\frac{(\mu_+ -x)(x- \mu_- )}{(\mu_+ -y)(y- \mu_- )}}
\int dz \rho(z) \big( f(y) + K(y-z) \big)
\cr
&=& \frac{1}{2\pi} \sum_{n=2}^\infty \sum_{k=0}^{n-2} \sum_{r=0}^{n-k-2}
 n g_n  \pi^{n/2} b_r b_{n-k-2-r} \mu_+^r \mu_-^{n-k-r-2} \sqrt{(\mu_+-x)(x-\mu_-)} x^k
\cr
&&-\frac{1}{\pi^2} 
\dashint_{\mu_-}^{\mu_+} \frac{dy}{x-y} \sqrt{\frac{(\mu_+ -x)(x- \mu_- )}{(\mu_+ -y)(y - \mu_- )}}
\int dz \rho(z) \big( K(y-z) - K(y) \big)
~. 
\eea
To obtain this result we used the identity \cite{Rodriguez-Gomez:2016ijh}
\beq
\label{freebja}
\dashint_{\mu_-}^{\mu_+} \frac{dy}{x-y} \frac{y^{n-1}}{\sqrt{(\mu_+-y)(y-\mu_-)}} 
= -\pi \sum_{k=0}^{n-2}\sum_{r=0}^{n-k-2} b_r b_{n-k-r-2} \mu_+^r \mu_-^{n-k-r-2} x^k
\eeq
with
\beq
\label{freebjb}
b_k = \frac{1}{\sqrt \pi} \frac{\Gamma\left( k+\frac{1}{2} \right)}{k!}
~.\eeq
Integrating this equation further over the domain of eigenvalues gives, in conjunction with the normalization 
condition \eqref{freebe},
\bea\label{freebk}
1&=& -\frac{1}{2} \sum_{n=2}^\infty \sum_{k=0}^{n-2} \sum_{r=0}^{n-k-2} \sum_{s=0}^{k+2}
n g_n \pi^{n/2} b_r b_{n-k-2-r} \sigma_s \sigma_{k-s+2} \mu_+^{r+s} \mu_-^{n-s}
\cr
&&- \frac{1}{\pi^2}  \dashint_{\mu_-}^{\mu_+} dx
\dashint_{\mu_-}^{\mu_+} \frac{dy}{x-y} \sqrt{\frac{(\mu_+ -x)(x- \mu_- )}{(\mu_+ -y)(y - \mu_- )}}
\int dz \rho(z) \big( K(y-z) - K(y) \big)
~.\eea
We used the identities \cite{Rodriguez-Gomez:2016ijh}
\beq
\label{freebka}
\int_{\mu_-}^{\mu_+} dx \sqrt{(\mu_+ - x) (x-\mu_-)} x^k =-\pi \sum_{s=0}^{k+2} \sigma_s \sigma_{k-s+2} \mu_+^s
\mu_-^{k-s+2}
~,\eeq
where
\beq
\label{freebkb}
\sigma_k =\frac{1}{2\sqrt \pi}\frac{\Gamma\left( k -\frac{1}{2}\right)}{k!}
~.\eeq

The simultaneous solution of the above equations determines $\mu_\pm$ and $\rho(x)$ parametrically in terms of the 
single-trace coupling constants $g_n$. Technical aspects of the solution are discussed in detail 
in the two subsequent appendices in different regimes. Once a solution is known the free energy is determined by the expression
\beq\label{freebl}
\FF(\{ g_n \}) = -\dashint dx\, dy \, \rho(x) \rho(y) \log \big( (x-y) H(x-y) \big) 
+ \int dx\, \rho(x)  \big( V(x) +2 \log H(x) \big)
~,\eeq
which can be manipulated further, using the saddle-point equations, to the more convenient form
\beq\label{freebm}
\FF(\{ g_n \}) = \int_{\mu_-}^{\mu_+} \rho(x) \left( \frac{1}{2} V(x) - \log|x| \right)
\eeq
that does not contain directly the special function $H(x)$.

It is useful to highlight here the following properties of the saddle-point equations and their solutions.

\vspace{0.5cm}
\noindent
{\bf A. Self-contained system of equations.}
There are obviously enough equations (\eqref{freebj}, \eqref{freebk}) 
to obtain the eigenvalue density $\rho(x)$ and the bounds of the support $\mu_\pm$ when 
$\mu_-=-\mu_+$. This occurs when the deformation is even, i.e.\ when only $g_n$ with $n$ even are present.
In that case the density is symmetric around the origin, $\rho(x)=\rho(-x)$. 
More generally, e.g.\ when odd interactions are allowed in the potential $V$, 
the eigenvalue density is not symmetric and $\mu_+\neq -\mu_-$. In that case, 
there is an additional non-trivial condition \cite{Rodriguez-Gomez:2016ijh,Marino:2011nm}
that can be used to fix the relation between
$\mu_+$ and $\mu_-$. It follows from the requirement that the resolvent 
\beq
\label{freebma}
\omega(x) = \dashint_{\mu_-}^{\mu_+} dy \frac{\rho(y)}{x-y} \sim \frac{1}{x}
\eeq
at large $x$. To ensure this drop-off the resolvent cannot contain any terms proportional to $x^n$ with $n>0$. 
This is equivalent to the requirement
\beq
\label{freebmb}
\dashint_{\mu_-}^{\mu_+} dx \frac{\omega(x)}{\sqrt{(\mu_+-x)(x-\mu_-)}} =0
~.\eeq
This equation is automatic when $\mu_+=-\mu_-$ and the density is even.

\vspace{0.5cm}
\noindent
{\bf B. Special cases vs generic $K$.}
When the function $K$ vanishes (this occurs in the case of $\NN=4$ SYM theory) the matrix integral has the standard 
measure and the exact solution of the saddle point equations follows immediately from \eqref{freebj}
\beq\label{freebn}
\rho(x) \big |_{K=0} =
 \frac{1}{2\pi} \sum_{n=2}^\infty \sum_{k=0}^{n-2} \sum_{r=0}^{n-k-2}
 n g_n \pi^{n/2}  b_r b_{n-k-2} \mu_+^r \mu_-^{n-k-r-2} \sqrt{(\mu_+-x)(x-\mu_-)} x^k
~.
\eeq
In the presence of a non-trivial function $K$, however, an exact solution is not known in closed form. In appendices
\ref{appweak} and \ref{appstrong} we analyze the density $\rho$ and free energy $\FF$ perturbatively in the strong and weak $g_2= \frac{8\pi^2}{\lambda}$
regime, respectively. Although it is possible to perform a comprehensive analysis of the weak-$\lambda$ regime at any desired order of 
perturbation theory, current techniques do not provide an equally satisfactory treatment of the strong coupling side.

\vspace{0.5cm}
\noindent
{\bf C. Universality.} Once a solution of the saddle-point equations is obtained, the result can be inserted into the free energy $\FF$ to obtain a generating 
function for the connected correlation functions of the model. Remarkably, the final expression of the correlation functions can be written solely 
in terms of $\mu_\pm$ (the limits of the eigenvalue support), and is independent of the details of the potential function $f$. We will give an example
of this property of universality in a moment for connected 2-point functions. 
The explicit dependence on the single-trace couplings that appear in the function $f$ arises from the 
corresponding dependence of the quantities $\mu_\pm$.

\subsection{Aside: density for connected 2-point functions of single-trace operators}
\label{2density}

In the previous subsection, and in appendices \ref{appweak}, \ref{appstrong}, we evaluate the free energy of the deformed matrix model at general values of the single-trace
couplings $g_n$ and take derivatives at $g_n=0$ for $n>2$, and $g_2=\frac{8\pi}{\lambda}$ free. This allows us to compute connected correlation functions
of the matrix model at any value of $g_2$ and $g_n=0$ for $n>2$. 

In this appendix we consider a different evaluation of the same connected correlation functions 
that does not go through the computation of the free energy in the deformed matrix model with $g_n\neq 0$ ($n>2$). We focus on the 2-point functions of 
single-trace operators. The result is expressed as a double integral over the eigenvalues with an appropriate 2-point function density $\bar\rho_2(x,y)$
\beq
\label{conDensaa}
\langle \Tr[\varphi^{n+1}] \Tr[\varphi^{m+1} ] \rangle_c = \int dx \int dy \, \bar\rho_2(x,y) \, x^{n+1} y^{m+1}
~.\eeq

There are well-known results in the literature regarding $\bar\rho_2$ (and its higher $n$-point generalizations) ---for a recent comprehensive review we refer the 
reader to \cite{Eynard:2015aea}. For example, in the case of the standard Hermitian matrix model
\beq
\label{conDensab}
\ZZ = \int dM \, e^{-N \, \Tr V(M)}
\eeq
where $V$ is a real polynomial of some degree, the leading large-$N$ contribution to the connected 2-point function density $\bar\rho_2$ takes the form
\beq
\label{condDensac}
\bar\rho_2 (x,y) = \frac{1}{2\pi^2} \frac{1}{(x-y)^2} \frac{xy-\mu^2}{\sqrt{x^2-\mu^2}\sqrt{y^2-\mu^2}}
~.
\eeq
We have assumed a single-cut solution of the saddle-point equations in the symmetric interval $[-\mu,\mu]$.
In accordance to point $C$ in the previous subsection, this expression is universal, namely independent of the details
of the potential $V$.

\subsubsection*{Integral equation for $\barM \rho_2$}

At this point we take the opportunity to derive a general integral equation obeyed by $\barM \rho_2$ (we are not
aware of a similar derivation in the literature). In an attempt to be fairly general, let us consider the large-$N$ limit 
of the matrix integral
\beq
\label{conDensad}
\ZZ_\OO = \int d^{N}a \prod_{i < j} \Delta^2 (a_i-a_j) \, e^{-N \sum_{i=1}^N V(a_i)} \prod_L \OO_L(a_i)
~,\eeq 
where $\OO_L$ are (local) single-trace operator insertions of the $N\times N$ Hermitian matrix over which we integrate. 
$\ZZ_{\bf 1}$, the matrix integral without any operator insertions, denotes the partition function of the matrix model.
By definition, the normalized correlation functions in this matrix model are of the form
\beq
\label{conDensae}
\langle \prod_L \OO_L \rangle = \frac{\ZZ_\OO}{\ZZ_1}
~.
\eeq
An elementary computation of $\ZZ_\OO$ in the saddle-point approximation leads to the following expressions.

For starters, the effective action for the eigenvalues $a_i$ is
\beq
\label{conDensaf}
S_{eff}(a_i) = \sum_{i=1}^N V(a_i) -\frac{1}{N} \sum_{i\neq j} \log \Delta (a_i-a_j)
-\frac{1}{N} \sum_L \log \OO_L (a_i)
~.\eeq
The {\it leading-order} saddle-point equations are
\bea
\label{conDensag}
\frac{\d V}{\d a_i} 
- \frac{2}{N} \sum_{k\neq i} ( \log \Delta)'(a_i - a_k) =0
~.\eea
We dropped the term 
\beq
-\frac{1}{N} \sum_L \d_{a_i} \log \OO_L 
\eeq
which is $\OO(1/N)$ (the other terms are $\OO(N^0)$). We denote the solution of the system of equations
\eqref{conDensag} as $a_i^{(0)}$. 

Then, setting (in standard fashion)
\beq
\label{conDensag2}
a_i = a_i^{(0)}+ \frac{1}{\sqrt N}\delta_i
\eeq
and performing the Gaussian integrations over $\delta_i$ we obtain 
\bea
\label{conDensai}
Z_\OO = 
\prod_{i<j} \Delta^2 \left(a_i^{(0)} - a_i^{(0)} \right)
e^{-N \sum_{i=1}^N  V(a_i^{(0)}) }
\prod_L \OO_L \left( a_i^{(0)} \right)
e^{\frac{1}{2} N P^T Q^{-1} P} e^{-\frac{1}{2} \log \frac{\det \, Q_\OO}{\det Q_{\bf 1}}}
~,\eea
where
$P$ and $Q_\OO$ are respectively the vector and matrix
\beq
\label{conDensaj}
P_i := \d_{a_i}S_{eff} \big |_0 = -\frac{1}{2}\sum_L \d_{a_i} \log \OO_L \big |_0~,
\eeq
\beq
\label{conDensak}
Q_{ij} := \d_{a_i a_j} S_{eff} \Big |_0 =
\frac{\d V}{\d a_i \d a_j}  \delta_{ij} 
-\frac{1}{N} \sum_L \d_{a_i} \d_{a_j} \log \OO_L  
-\frac{1}{N} \sum_{k\neq m} \d_{a_i} \d_{a_j} \log \Delta (a_m -a_k) \Big |_{a^{(0)}}
~.\eeq
After a few straightforward algebraic manipulations one can show that up to order $\OO(N^{-2})$
\beq
\label{conDensal}
\langle \prod_L \OO_L \rangle = \frac{Z_\OO}{Z_{\bf 1}}
= \prod_L \OO_L \left( a_I^{(0)} \right) 
\left( 1 + \frac{1}{2} N P^T Q^{-1} P - \frac{1}{2} \Tr \left[ Q^{-1} dP \right] 
+\OO(N^{-4}) \right)
~.
\eeq
$dP$ is the matrix with elements 
\beq
\label{conDensam}
\left( dP \right)_{ij} = -\frac{1}{N} \sum_{L} \d_{a_i} \d_{a_j} \log \OO_L \Big |_0
~.
\eeq
On the r.h.s.\ of equation \eqref{conDensal} the first term is the factorizable disconnected part of the correlation function
and the second part is the first subleading contribution in $1/N$.

Now assume that the saddle-point configuration is a one-cut solution on the symmetric interval $[-\mu,\mu]$
with eigenvalue density $\rho(x)$, and consider the two-point function of two single-trace operators, namely set
\beq
\label{conDensan}
\prod_L \OO_L = \Tr[\varphi^{n+1}]\Tr[\varphi^{m+1}]
~.\eeq
Moreover, for concreteness, let us specialize further to the matrix model of interest in this paper where
\beq
\label{conDensana}
\Delta(x) = |x| H(x)~~, ~~ V(x) = g_2 x^2 + 2 \log H(x)
~.\eeq

Then, in the continuum limit explicit evaluation gives
\bea
\label{conDensao}
&&\frac{1}{2} N\, P^T Q^{-1} P-\frac{1}{2} \Tr \left[ Q^{-1} d P \right]
= \frac{1}{N}  \int_{-\mu}^\mu dx \int_{-\mu}^\mu dy\, \rho(x) \rho(y) Q^{-1}(x,y)
\frac{(n+1)(m+1)}{m_{n+1}m_{m+1}} x^n y^{m}
\cr
&&+\frac{1}{2N} \int_{-\mu}^\mu dx \, \rho(x) Q^{-1}(x,x) 
\left[ \frac{n(n+1)}{m_{n+1}} x^{n-1} 
+ \frac{m (m+1)}{m_{m+1}} x^{m-1} \right]
~,\eea
where $m_n$ are the moments
\beq
\label{conDensap}
m_n = \int_{-\mu}^\mu dx \, \rho(x) x^n
~,\eeq
and $Q^{-1}(x,y)$ is the functional inverse of $Q(x,y)$
\beq
\label{conDensaq}
\dashint_{-\mu}^\mu dz \, \rho(z) Q(x,z) Q^{-1}(z,y) = \frac{1}{N^2} \frac{\delta(x-y)}{\rho(y)}
~.\eeq
Combining this equation with the definition of $Q$ \eqref{conDensak} (in its continuous form) we deduce that the
leading $\OO(N^{-1})$ part of $Q^{-1}$ obeys the more explicit integral equation
\beq
\label{conDensar}
\dashint_{-\mu}^\mu dz\, \rho(z) \left( \frac{1}{(x-z)^2}+K'(x-z) \right) Q^{-1}(z,y) = \frac{1}{2N} \frac{\delta(x-y)}{\rho(y)}
~.\eeq

Equation \eqref{conDensao} has two characteristic features:
\begin{itemize}
\item[$(i)$] The first term on the r.h.s. depends only on the combination
\beq
\label{conDensas}
F_2(x,y) := \rho(x) \rho(y) Q^{-1}(x,y)
~.\eeq
\item[$(ii)$] The second term on the r.h.s.\ involves the singular quantity $Q^{-1}(x,x)$. Part of the prescription we 
propose to adopt in the evaluation of the above expressions is to remove this singular term by hand. We will soon see
that this prescription works quite well and in agreement with known results in the standard Hermitian matrix model,
which corresponds to the special case $K=0$.
\end{itemize}

Consequently, with these specifications the connected two-point function 
$\langle \Tr[\varphi^{n+1}]\Tr[\varphi^{m+1}]\rangle_c$ that follows from \eqref{conDensal} takes the form
\beq
\label{conDensat}
\langle \Tr[\varphi^{n+1}]\Tr[\varphi^{m+1}]\rangle_c = 
\frac{1}{N} \dashint_{-\mu}^\mu dx \dashint_{-\mu}^\mu dy \, F_2(x,y) \d_x (x^{n+1}) \d_y (y^{m+1})
~.\eeq
Requiring the boundary conditions
\beq
\label{conDensau}
F_2(\pm \mu,y)=0~, ~~ \d_x F_2(x,\pm \mu)=0
\eeq
we can perform two integrations by part to recast \eqref{conDensat} as
\beq
\label{conDensav}
\langle \Tr[\varphi^{n+1}]\Tr[\varphi^{m+1}]\rangle_c
= \frac{1}{N} \dashint_{-\mu}^\mu dx \dashint_{-\mu}^\mu dy \, \d_x \d_y F_2(x,y) x^{n+1}\, y^{m+1}
~.\eeq
This implies
\beq
\label{conDensba}
\bar\rho_2(x,y) = N  \d_x \d_y F_2(x,y)
~.\eeq
We will discuss the self-consistency of the boundary conditions \eqref{conDensau} in a moment.

Armed with the relation \eqref{conDensba} and the equation \eqref{conDensar} obeyed by the inverse $Q^{-1}(x,y)$
we are now in position to formulate an integral equation for the density of connected 2-point functions $\bar\rho_2$.
Integrating by parts on the integral of the l.h.s.\ in eq.\ \eqref{conDensar} and using the first of the boundary conditions
\eqref{conDensau} we obtain 
\beq
\label{conDensbb}
\dashint_{-\mu}^\mu dz \left( \frac{1}{x-z} - K(x-z) \right) \d_z F_2 (z,y) 
= -\frac{1}{2N} \delta(x-y)
~.\eeq
Applying the integral
\beq
\label{conDensbc}
\dashint_{-\mu}^\mu dx \sqrt{\mu^2 -x^2} \frac{1}{w-x}
\eeq
on both sides of \eqref{conDensbb} we obtain 
\beq
\label{conDensbd}
\d_w F_2(w,y) = -\frac{1}{2N\pi^2}  \sqrt{\frac{\mu^2-y^2}{\mu^2-w^2}} \frac{1}{w-y}
-\frac{1}{\pi^2} \dashint_{-\mu}^\mu dx 
\sqrt{\frac{\mu^2-x^2}{\mu^2-w^2}} \frac{1}{w-x} 
\int dz\, K(x-z)  \d_z F_2(z,y)
~,
\eeq
which is an integral equation for the derivative of $F_2$, $\d_x F_2(x,y)$. Solving this equation, or its progenitor
\eqref{conDensbb}, and applying a second derivative on the second argument of $F_2$ we obtain the density
$\barM \rho_2$ that allows us to determine the connected two-point function of single-trace operators at any value of  
the 't Hooft coupling at leading order in the $1/N$ expansion.

As an additional comment on the boundary conditions \eqref{conDensau}, notice that the first one is a natural 
consequence of the assumption that the inverse $Q^{-1}(x,y)$ is regular at $x=\pm \mu$ (or not singular enough ---an
example is provided by the standard Hermitian matrix model, $K=0$, below) and the fact that the density of eigenvalues
vanishes at the boundaries of the eigenvalue support, $\rho(\pm \mu)=0$. The second boundary condition in 
\eqref{conDensau} is a self-consistent ansatz from the point of view of the integral equation \eqref{conDensbd}.

\subsubsection*{Recovering the standard Hermitian matrix model formula}

As a test of the proposed relation \eqref{conDensba} consider the case of the Hermitian matrix model \eqref{conDensab},
which corresponds to the special case $K=0$ in the above analysis. Then, a derivative of eq.\ \eqref{conDensbd}, where
only the first term on the r.h.s\ contributes, gives
\beq
\label{conDensca}
\bar\rho_2(x,y) = N \d_x\d_y F_2(x,y) = \frac{1}{2\pi^2} \frac{1}{(x-y)^2} \frac{xy-\mu^2}{\sqrt{x^2-\mu^2}\sqrt{y^2-\mu^2}}
\eeq
in exact agreement with the known result \eqref{condDensac}.

\subsubsection*{$\barM \rho_2$ at infinite 't Hooft coupling in the matrix model of $SU(N)$ $\NN=2$ SCQCD}

An analytic solution of eq.\ \eqref{conDensbb}, or \eqref{conDensbd}, at all values of the 't Hooft coupling $\lambda$
is currently not known. It is straightforward, however, to obtain the analytic solution at infinite coupling. Since, the 
strong coupling regime is hard to analyze with existing methods, it is of some value to report here the analytic
profile of $\bar\rho_2$ at infinite coupling. 

For this purpose it is convenient to return to eq.\ \eqref{conDensbb}. Since $\d_x F_2(x,y)$ is a function of the
difference $x-y$, we can perform a simple Fourier transformation
\bea
\label{conDensda}
&&\d_x F_2 (x) =\frac{1}{\sqrt{2\pi}} \int_{-\infty}^\infty dk \, e^{i k x} \widehat {\d_x F_2}(k)
\cr
&&\widehat {\d_x F_2}(k) =\frac{1}{\sqrt{2\pi}} \int_{-\infty}^\infty dx \, e^{- i k x} \d_x F_2(x)
\eea
to obtain
\beq
\label{conDensdb}
\widehat{\d_x F_2}(k) = \frac{i}{N\pi} {\rm sgn}(k) \frac{\sinh^2 \frac{k}{2}}{\cosh k}
~,\eeq
which gives in real space
\bea
\label{conDensdc}
\barM \rho_2(x-y)\big |_{\lambda=\infty} &=& \frac{1}{16 \sqrt{2} N \pi^{3/2}}
\bigg[ \frac{16}{(x-y)^2}
+ \psi' \left( \frac{1}{4}(1+i(x-y))\right) 
+  \psi' \left( \frac{1}{4}(1-i(x-y))\right) 
\cr
&&+ \psi' \left( \frac{1}{4}(3+i(x-y))\right)
+ \psi' \left( \frac{1}{4}(3-i(x-y))\right)  
\bigg]
~.\eea
As usual, $\psi(z)=\frac{\Gamma'(z)}{\Gamma(z)}$ is the logarithmic derivative of the $\Gamma$-function.

Unfortunately, this datum is not enough to determine the large-$\lambda$ behavior of the connected single-trace
two-point functions. When we apply the formula \eqref{conDensaa} at large but finite $\lambda$ we need to know
also the first subleading (in $\frac{1}{\lambda}$) correction to $\bar\rho_2$. The mere knowledge of 
\eqref{conDensdc} appears to provide the correct leading large-$\mu$ scaling of the connected two-point functions 
but fails to capture the exact numerical coefficient.

\subsubsection*{Towards higher connected $n$-point functions}

Working directly with the densities of connecting $n$-point functions (as we did above for $n=2$) is a direction 
with a potential for promising results. In the past considerable results have been obtained in standard matrix
models through the analysis of the matrix model loop equations which provide recursive equations between
the generating functions of connected $n$-point functions (see \cite{Eynard:2015aea} for a modern review
and references to the original literature). Such results can be extended to more general 
cases (like \eqref{conDensad}) ---see, for instance, the recent work \cite{Borot:2013lpa}. 
It would be interesting to pursue
this approach for the matrix models of interest in this paper. We hope to return to this aspect in a future publication.

\section{Proof of \eqref{3ptflatsphere}}
\label{proof}

In this appendix we prove formula \eqref{3ptflatsphere}. Let us first recall that the 3-point function we are interested in is given by the formula
\beq
\label{eqn:appfull3pt}
\left<\left(O_{k_1}O_{k_2}\right)^{\IR^4} , O^{\mathbb{R}^4}_{k_3}\right>~,
\eeq
where, as always, the $\IR^4$ indices on the operators indicate that we have applied the \emph{full} Gram-Schmidt
procedure, i.e.\ the primed operators are orthogonal to all the operators of lower conformal dimension, both single- and
multi-trace. We want to prove that in the large-$N$ limit the 2-point function above is given by the simpler expression
\beq
\label{eqn:appst3pt}
\left<\left(O^{\mathbb{R}^4}_{k_1}O^{\mathbb{R}^4}_{k_2}\right) \, ,O^{\mathbb{R}^4}_{k_3}\right>~,
\eeq
where the operators $O^{\mathbb{R}^4}_{k}$ are orthogonal to all the single-trace operators of lower conformal dimension
\emph{only}, and $\left(O^{\mathbb{R}^4}_{k_1}O^{\mathbb{R}^4}_{k_2}\right)$ is simply the product of the operators
$O^{\mathbb{R}^4}_{k_1}$ and $O^{\mathbb{R}^4}_{k_2}$.

We first examine the difference between the operators $\left(O_{k_1}O_{k_2}\right)^{\IR^4}$ and $\left(O^{\mathbb{R}^4}_{k_1}O^{\mathbb{R}^4}_{k_2}\right)$. We have
\beq
\left( O_{k_1}O_{k_2}\right)^{\IR^4} = O^{S^4}_{k_1}O^{S^4}_{k_2} + (\textrm{single-trace}) + \sum_{\ell_1\leq\ell_2} c_{k_1,k_2}^{\ell_1,\ell_2} \left(O_{\ell_1}O_{\ell_2}\right)^{\IR^4} + \frac{1}{N}(\textrm{higher traces})~,
\eeq
where the last three terms on the r.h.s.\ involve operators of conformal dimension less than $k_1 + k_2$. Let us analyze
the double-trace coefficients $c_{k_1,k_2}^{\ell_1,\ell_2}$. They are given by
\beq
c_{k_1,k_2}^{\ell_1,\ell_2} = \frac{\left<O^{S^4}_{k_1}O^{S^4}_{k_2}\, , \left(O_{\ell_1}O_{\ell_2}\right)^{\IR^4}\right>}{\left<\left(O_{\ell_1}O_{\ell_2}\right)^{\IR^4} , \left(O_{\ell_1}O_{\ell_2}\right)^{\IR^4} \right>}~.
\eeq
Using large $N$-factorization, we obtain
\beq
\sum_{\ell_1\leq\ell_2} c_{k_1,k_2}^{\ell_1,\ell_2} \left(O_{\ell_1}O_{\ell_2}\right)^{\IR^4} 
= \sum_{\ell_1,\ell_2} c_{k_1}^{\ell_1} c_{k_2}^{\ell_2}\, O^{\mathbb{R}^4}_{\ell_1}O^{\mathbb{R}^4}_{\ell_2} + \OO(1/N^2)
\eeq
where  $c_{k}^{\ell}$ is given by
\beq
c_{k}^{\ell} = \frac{\left<O^{S^4}_{k}, O^{\mathbb{R}^4}_{\ell}\right>}{\left<O^{\mathbb{R}^4}_{\ell},O^{\mathbb{R}^4}_{\ell}\right>}~.
\eeq
From this we conclude that
\beq
\label{eqn:mtdec}
\left(O_{k_1}O_{k_2}\right)^{\IR^4} = \left(O^{\mathbb{R}^4}_{k_1}O^{\mathbb{R}^4}_{k_2}\right) + (\textrm{single-trace}) + \frac{1}{N}(\textrm{higher traces}) + \OO(1/N^2)~.
\eeq
Also recall that up to $\OO(1/N)$ corrections the linear combinations $\OO_k^{\IR^4}$ involve only single-trace
operators.

If we now examine again equation \eqref{eqn:appfull3pt}, we see that the single-trace operators on the r.h.s.\ of
\eqref{eqn:mtdec} do not contribute since $O^{\mathbb{R}^4}_{k_3}$ is orthogonal to all of them, while the higher-trace
operators are suppressed in the large-$N$ limit. This proves \eqref{3ptflatsphere}. Notice that it is important that the
operator $\left(O^{\mathbb{R}^4}_{k_1}O^{\mathbb{R}^4}_{k_2}\right)$ does not appear in the r.h.s.\ of the definition of
$O_{k_3}^{\IR^4}$, otherwise it could give a contribution of the same order as \eqref{eqn:appst3pt} by large-$N$
factorization. For correlators that respect $R$-charge conservation, namely $k_3 = k_1 + k_2$, the higher-trace
operators in the full definition of $O_{k_3}^{\IR^4}$ must have conformal dimension $\Delta < k_1 + k_2$, so
$\left(O^{\mathbb{R}^4}_{k_1}O^{\mathbb{R}^4}_{k_2}\right)$ indeed cannot appear. In the case $k_3 > k_1 + k_2$,
applying \eqref{3ptflatsphere} would give a non-zero result, inconsistent with $R$-charge conservation. However, this is
precisely the case where we cannot ignore the double-trace operator
$\left(O^{\mathbb{R}^4}_{k_1}O^{\mathbb{R}^4}_{k_2}\right)$ in the expansion of $O_{k_3}^{\IR^4}$; its effect is to
precisely cancel the non-zero contribution coming from \eqref{3ptflatsphere}.

In summary, we have proven that the formula \eqref{3ptflatsphere} gives the correct 3-point function in the large-$N$
limit for operators that respect $R$-charge conservation, which are the only non-zero 3-point functions on the plane.

\section{Large-$N$ matrix model at weak coupling}
\label{appweak}
In this short appendix we collect some explicit expressions for the three-point functions at tree-level. Similar results
were presented in \cite{Rodriguez-Gomez:2016ijh}. Here we generalize the results of appendix C in
\cite{Rodriguez-Gomez:2016ijh} to obtain three-point functions for single-trace operators of both even and odd powers at
tree-level. This serves as input for the closed form diagonalization procedure discussed in section \ref{weakcoupling}.

As reviewed in appendix \ref{deformedmatrix} (see equations \eqref{freebf},\eqref{freebg},\eqref{freebi}) in the
continuum limit we obtain the saddle-point equation
\begin{equation}
\frac{1}{2}\sum_{n=1}g_{n}\pi^{\frac{n}{2}}nx^{n-1}-K(x)=\dashint_{-\mu_-}^{\mu_+} dy \, \rho(y)\left(\frac{1}{x-y}-K(x-y)\right).
\end{equation}
This equation can be inverted to get an expression for $\rho(x)$ by applying the integral operator
$\dashint_{-\mu_-}^{\mu_+}\frac{dx}{\sqrt{(\mu_+-x)(\mu_-+x)}(x-z)}$. Further use of the integral identities
\eqref{freebja}, \eqref{freebka} provides implicit relations for the moments up to an arbitrary but finite number of
loops
\begin{multline}
m_{q}=-\frac{1}{2}\sum_{n=1}ng_{n}\pi^{\frac{n}{2}}\sum_{k=0}^{n-2}\sum_{r=0}^{n-k-2}b_{r}b_{n-k-r-2}\sum_{s=0}^{q+k+2}\sigma_{s}\sigma_{k+q-s+2}\mu_{+}^{r+s}\mu_-^{n+q-r-s}\\+2\sum_{p=1}^{M}(-1)^{p}\zeta(2p+1)\sum_{k=1}^{2p}(-1)^{k}\binom{2p+1}{k}m_{k}\sum_{l=0}^{2p-k}\sum_{r=0}^{2p-k-l}b_{r}b_{2p-k-r-l}\\\times\sum_{s=0}^{q+l+2}\sigma_{s}\sigma_{q+l+2-s}\mu_{+}^{r+s} \mu_-^{2p+q+2-k-r-s}.
\label{weakmoments}
\end{multline}
The moments appear linearly in this expression, hence by truncating the perturbation theory to a finite order 
and solving the corresponding system of linear equations 
we obtain expressions for the moments accurate up to that order. 

In what follows it will be useful to define a separate expression for the first line of the previous expression
\begin{equation}
\phi_{q}(g_n)=-\frac{1}{2}\sum_{n=1}n\pi^{\frac{n}{2}}g_{n}\sum_{k=0}^{n-2}\sum_{p=0}^{n-k-2}b_{p}b_{n-k-p-2}\sum_{s=0}^{k+q+2}\mu_{+}^{p+s} \mu_-^{n+q-p-s}\sigma_{k+q-s+2}\sigma_{s}~.
\end{equation}
The function $\phi_{q}(g_n)$ is equal to the tree-level contribution to the moments. For quick reference, we note here
its second derivative with all couplings set to zero, except $g_{2}$ (and hence appropriately $\mu_{-}$ set to
$-\mu_{+}$)
\begin{multline}
\frac{\partial^2\phi_q}{\partial g_{a}\partial g_{c}}=-\frac{1}{2}\frac{\partial\mu_+}{\partial g_{c}}a\pi^{\frac{a}{2}}\sum_{k=0}^{a-2}\sum_{p=0}^{a-k-2}b_{p}b_{a-k-p-2}\sum_{s=0}^{k+q+2}(p+s)(-1)^{a+q-p-s}\mu_{+}^{a+q-1}\sigma_{k+q-s+2}\sigma_{s}\\-\frac{1}{2}\frac{\partial\mu_+}{\partial g_{a}}c\pi^{\frac{c}{2}}\sum_{k=0}^{c-2}\sum_{p=0}^{c-k-2}b_{p}b_{c-k-p-2}\sum_{s=0}^{k+q+2}(p+s)(-1)^{c+q-p-s}\mu_+^{c+q-1}\sigma_{k+q-s+2}\sigma_{s}\\+\frac{1}{2}\frac{\partial\mu_-}{\partial g_{c}}a\pi^{\frac{a}{2}}\sum_{k=0}^{a-2}\sum_{p=0}^{a-k-2}b_{p}b_{a-k-p-2}\sum_{s=0}^{k+q+2}(a+q-p-s)(-1)^{a+q-p-s-1}\mu_{+}^{a+q-1}\sigma_{k+q-s+2}\sigma_{s}\\+\frac{1}{2}\frac{\partial\mu_-}{\partial g_{a}}c\pi^{\frac{c}{2}}\sum_{k=0}^{c-2}\sum_{p=0}^{c-k-2}b_{p}b_{c-k-p-2}\sum_{s=0}^{k+q+2}(c+q-p-s)(-1)^{c+q-p-s}\mu_{+}^{c+q-1}\sigma_{k+q-s+2}\sigma_{s}\\+\left(\frac{\partial\mu_{+}}{\partial g_{a}}\frac{\partial\mu_-}{\partial g_{c}}+\frac{\partial \mu_-}{\partial g_{a}}\frac{\partial\mu_+}{\partial g_{c}}\right)\pi g_{2}\sum_{s=0}^{q+2}s(q-s+2)(-1)^{q-s+1}\mu_{+}^{q}\sigma_{q-s+2}\sigma_{s}\\-\pi g_{2}\sum_{s=0}^{q+2}\sigma_{s}\sigma_{q-s+2}s(-1)^{q-s}\left((s-1)\frac{\partial\mu_{+}}{\partial g_{a}}\frac{\partial\mu_{+}}{\partial g_{c}}+\mu_{+}\frac{\partial^2\mu_{+}}{\partial g_{a}\partial g_{c}}\right)\mu_{+}^{q}\\-\pi g_{2}\sum_{s=0}^{q+2}\sigma_{s}\sigma_{q-s+2}(q-s+2)(-1)^{q-s}\left((q-s+1)\frac{\partial\mu_-}{\partial g_{a}}\frac{\partial\mu_-}{\partial g_{c}}+\mu_{+}\frac{\partial^2\mu_-}{\partial g_{a}\partial g_{c}}\right)\mu_{+}^{q}~.\\
\label{2ndderivmom}
\end{multline}

\subsection{Results at tree-level}
At tree-level the moments simply reduce to $m_{q}=\phi_{q}(g_n)$. In addition, when all the couplings except $g_2$ have
been set to zero the endpoints of the eigenvalue distribution are given by the expressions
$\mu_+(g_2)=-\mu_-(g_2)=\sqrt{\frac{2}{\pi g_2}}$. To specify the derivatives of the endpoints with respect to the
coupling $g_{n}$ we require two conditions. The first condition comes from the normalization of the eigenvalue density
\begin{equation}
m_0=\int_{-\mu_{-}}^{\mu_+}\rho(x)dx=1.
\label{condition1}
\end{equation}
The second condition, \eqref{freebmb}, turns into the constraint
\begin{equation}
0=\frac{\pi}{2}\sum_{n=1}n g_{n}\pi^{\frac{n}{2}}\sum_{k=0}^{n-1}b_{k}b_{n-k-1}\mu_{+}^{k} \mu_-^{n-k-1}
~.\label{condition2}
\end{equation}
By taking implicit derivatives of the conditions (\ref{condition1}) and (\ref{condition2}) we can determine the
derivatives of the endpoints by means of Gaussian elimination
\begin{multline}
\frac{\partial\mu_{+}}{\partial g_{a}}=-\frac{a\pi^{\frac{a}{2}}}{2\sqrt{2\pi g_{2}}}\sum_{k=0}^{a-2}\sum_{p=0}^{a-k-2}b_{p}b_{a-k-p-2}\sum_{s=0}^{k+2}\sigma_{s}\sigma_{k-s+2}(-1)^{a-p-s}\left(\frac{2}{\pi g_{2}}\right)^{\frac{a}{2}}\\-\frac{a\pi^{\frac{a}{2}}}{2\pi g_{2}}\sum_{k=0}^{a-1}b_{k}b_{a-k-1}(-1)^{a-k-1}\left(\frac{2}{\pi g_{2}}\right)^{\frac{a-1}{2}},
\end{multline}
and
\begin{multline}
\frac{\partial\mu_-}{\partial g_{a}}=-\frac{a\pi^{\frac{a}{2}}}{2\sqrt{2\pi g_2}}\sum_{k=0}^{a-2}\sum_{p=0}^{a-k-2}b_{p}b_{a-k-p-2}\sum_{s=0}^{k+2}\sigma_{s}\sigma_{k-s+2}(-1)^{a-p-s}\left(\frac{2}{\pi g_2}\right)^{\frac{a}{2}}\\+\frac{a\pi^{\frac{a}{2}}}{2\pi\tau}\sum_{k=0}^{a-1}b_{k}b_{a-k-1}(-1)^{a-k-1}\left(\frac{2}{\pi g_2}\right)^{\frac{a-1}{2}}.
\end{multline}
By means of quadratic transformations these expressions can be further simplified to
\begin{equation}
\frac{\partial\mu_+}{\partial g_{a}}=-\frac{a\pi^{\frac{a}{2}}}{4\pi g_2}((-1)^{a-1}+1)\left(\frac{2}{\pi g_2}\right)^{\frac{a-1}{2}}b_{\frac{a-1}{2}}-\frac{a\pi^{\frac{a}{2}}}{4\sqrt{2\pi g_2}}((-1)^a+1)\left(\frac{2}{\pi g_2}\right)^{\frac{a}{2}}b_{\frac{a}{2}},
\end{equation}
and
\begin{equation}
\frac{\partial\mu_-}{\partial g_{a}}=\frac{a\pi^{\frac{a}{2}}}{4\pi g_2}\left((-1)^{a-1}+1\right)\left(\frac{2}{\pi g_2}\right)^{\frac{a-1}{2}}b_{\frac{a-1}{2}}-\frac{a\pi^{\frac{a}{2}}}{4\sqrt{2\pi g_2}}((-1)^a+1)\left(\frac{2}{\pi g_2}\right)^{\frac{a}{2}}b_{\frac{a}{2}}.
\end{equation}

Similarly, by taking an additional derivative of the conditions we can find expressions for the second derivatives
\begin{equation}
\frac{\partial^2\mu_{+}}{\partial g_{a}\partial g_{c}}=-\frac{f(g_2,a,c)}{\pi^2 g_{2}}-\frac{h(g_2,a,c)}{\sqrt{2\pi g_{2}}}~,
\end{equation}
and
\begin{equation}
\frac{\partial^2\mu_{-}}{\partial g_{a}\partial g_{c}}=\frac{f(g_2,a,c)}{\pi^2 g_{2}}-\frac{h(g_2,a,c)}{\sqrt{2\pi g_{2}}}~,
\end{equation}
where respectively
\begin{multline}
f(g_{2},a,c)=\frac{1}{2}a\pi^{\frac{a+2}{2}}(-1)^{a}\left(\frac{2}{\pi g_2}\right)^{\frac{a-2}{2}} \times
\nonumber\\
\left[\left(\frac{\partial\mu_{+}}{\partial g_c}+\frac{\partial\mu_{-}}{\partial g_c}\right)\sigma_{a-2} \,_{2}F_{1}\left(\frac{3}{2},2-a;\frac{5}{2}-a;-1\right)+\frac{1}{2}\frac{\partial\mu_{-}}{\partial g_c}((-1)^{a-1}+1)b_{\frac{a-1}{2}}\right]
+(a\leftrightarrow c)~,
\end{multline}
and
\begin{multline}
h(g_{2},a,c)=-\frac{1}{2}a\pi^{\frac{a}{2}}\frac{\partial\mu_{+}}{\partial g_{c}}\sum_{k=0}^{a-2}\sum_{p=0}^{a-k-2}b_{p}b_{a-k-p-2}\sum_{s=0}^{k+2}(p+s)(-1)^{a-p-s}\left(\frac{2}{\pi g_{2}}\right)^{\frac{a-1}{2}}\sigma_{k-s+2}\sigma_{s}\\+\frac{1}{2}a\pi^{\frac{a}{2}}\frac{\partial\mu_-}{\partial g_{c}}\sum_{k=0}^{a-2}\sum_{p=0}^{a-k-2}b_{p}b_{a-k-p-2}\sum_{s=0}^{k+2}(a-p-s)(-1)^{a-p-s-1}\left(\frac{2}{\pi g_{2}}\right)^{\frac{a-1}{2}}\sigma_{k-s+2}\sigma_{s}\\\frac{1}{8}\pi g_{2}\left(\frac{\partial\mu_{+}}{\partial g_{a}}\frac{\partial\mu_{+}}{\partial g_{c}}+\frac{\partial\mu_{-}}{\partial g_{a}}\frac{\partial\mu_{-}}{\partial g_{c}}+\frac{\partial\mu_{+}}{\partial g_{a}}\frac{\partial\mu_{-}}{\partial g_{c}}+\frac{\partial\mu_-}{\partial g_{a}}\frac{\partial\mu_{+}}{\partial g_{c}}\right)+(a\leftrightarrow c)~.
\end{multline}

Substituting these expression for the derivatives of the endpoints back into expression (\ref{2ndderivmom}) gives us an
explicit expression for the second derivative of the moments which can be related back to the tree-level three-point
function of single-trace operators on the sphere by
\begin{equation}
\langle \Tr\left[ \varphi^{a} \right]\Tr\left[ \varphi^{c} \right]\Tr\left[ \varphi^{q} \right]\rangle_{S^4}=\pi^{\frac{q}{2}}\frac{\partial^2\phi_q}{\partial g_{a}\partial g_{c}}~.
\end{equation}
The combination of these results with the formulae developed in section \ref{23higher} yields results on the tree-level
flat-space three-point functions and their first subleading correction.

\section{Large-$N$ matrix model at strong coupling}
\label{appstrong}
The analysis of \cite{Passerini:2011fe} at infinite coupling shows that the size of the cut 
of the eigenvalue distribution grows with the value of the 't Hooft coupling. 
In that case, a more sophisticated treatment of the saddle-point equations is needed. Ref.\ 
\cite{Passerini:2011fe} proposed an approximate Wiener-Hopf approach.
Similar to the weak coupling case we can extend this method by adding appropriate polynomial sources to the free
energy. In this appendix we consider only the case of even source terms.
This extension was also suggested and partially implemented in \cite{Rodriguez-Gomez:2016ijh}.

For the benefit of the reader, and in order to set up the appropriate notation, we begin with a quick review of the
Wiener-Hopf method used in \cite{Passerini:2011fe}.

\subsection{Wiener-Hopf method}

Employing the following integral identity for the function $K(x)$
\begin{equation}
K(x)=\int_{-\infty}^{\infty}dw\, \frac{w \coth(\pi w)}{x-w}
\end{equation}
we can rewrite our saddle-point equation as a slightly generalized version of equation (4.15) of \cite{Passerini:2011fe}
\begin{multline}
\rho(x)-\int dy\, \rho(y)(x-y)\coth\pi(x-y)+x\coth(\pi x)=\\\frac{1}{2\pi}\sqrt{\mu^2-x^2}\sum_{n=1}n g_{2n}\pi^{n}\sum_{p=0}^{n-1}\frac{2n-2p-3}{n-n-1}\mu^{2n-2p-2}b_{n-p-2}x^{2p}\\-\frac{1}{\pi}\int_{|w|>\mu}\frac{dw}{w-x}\sqrt{\frac{\mu^2-x^2}{w^2-\mu^2}}\int dy\, \rho(y)\left[(w-y)\coth\pi(x-y)-w\coth(\pi w)\right],
\label{strongsaddle}
\end{multline}
where the coefficients $b_{k}$ are given in \eqref{freebjb}. As an integral equation this expression strictly holds true
only within the interval $[-\mu,\mu]$. This means that as an operator the integration kernel is singular. In the
Wiener-Hopf method we can exploit the knowledge that $\rho(x)$ is zero outside of the interval. Accordingly, the goal is
to find an expression whose inverse Fourier transfrom vanishes outside of the interval. A general method for this was
given in \cite{Novokshenov}, but it requires solving a very non-trivial factorization problem. We will not attempt to
solve this factorization problem here. Instead, following \cite{Passerini:2011fe}, we will aim to find a function that
is guaranteed to vanish only for $x>\mu$. We will then argue that as long as $\mu \gg 1$ the other endpoint will not be
of major concern. As a result, we need to find an expression of the form
\begin{equation}
\rho(k)=e^{i\mu k}f(k)~,
\label{FourierExample}
\end{equation}
where $f(k)$ is a function that is analytic in the lower half-plane. Computing the inverse Fourier transform by means of
contour integration shows us that this satisfies the boundary condition.

To find an expression of the form \eqref{FourierExample} we proceed by first computing the Fourier transform of equation (\ref{strongsaddle}):
\begin{equation}
\int_{-\infty}^{\infty}\frac{dk}{2\pi}e^{-ikx}\left(\frac{\cosh(k)}{2\sinh^2(\frac{k}{2})}\rho(k)-\frac{1}{2\sinh^2(\frac{k}{2})}-F(k)+e^{-i\mu k}X_-(k)+e^{i\mu k}X_+(k)\right)=0~,
\label{saddlepointstrong}
\end{equation}
where $F(k)$ is the Fourier transform of the right-hand side of (\ref{strongsaddle}), and where everything that is not
strictly determined by the integral equation is encapsulated by the two free functions $X_-(k)$ and $X_+(k)$. Note that
if $X_{\pm}(k)$ is analytic in, respectively, the upper or lower half-plane then $X_{\pm}(x)=0$ for either $x>\mu$ or
$x<-\mu$.

The key step in what follows is that the integration kernel in Fourier space has a known analytic decomposition
\begin{equation}
\frac{\cosh(k)}{2\sinh^2\left(\frac{k}{2}\right)}=\frac{1}{G_-(k)G_+(k)}
\end{equation}
where
\begin{equation}
G_{\pm}(k)=\frac{\sqrt{8\pi^3}2^{\pm \frac{ik}{\pi}}\Gamma\left(\frac{1}{2}\mp\frac{ik}{\pi}\right)}{k\Gamma^2\left(\mp\frac{ik}{2\pi}\right)}~.
\end{equation}
Most notably, these functions are respectively analytic in either the upper or lower half-plane and they go to zero
sufficiently fast as $k\rightarrow\infty$. Also note that their poles are located at respectively $\mp i\nu_{l}$ where
$\nu_l=\pi(l+\frac{1}{2})$. At these poles the residues are given by
\begin{equation}
r_{l}=\frac{(-2)^{l+1}\Gamma^2\left(\frac{l}{2}+\frac{5}{4}\right)}{\sqrt{\pi}\left(l+\frac{1}{2}\right)\Gamma(l+1)}
~.
\end{equation}

Next we multiply the integrand of (\ref{saddlepointstrong}) with $G_{+}(k)$ and subtract the poles in the lower
half-plane by means of the integral transform
\begin{equation}
\mathcal{F}_{-}(k)=-\int_{-\infty}^{\infty}\frac{dk'}{2\pi}\frac{\mathcal{F}(k')}{k'-k+i\epsilon}.
\end{equation}
This operation provides the expression
\begin{multline}
\label{rhoequation}
\frac{\rho(k)e^{-i\mu\omega}}{G_{-}(k)}=-\int^{\infty}_{-\infty}\frac{dk'}{2\pi i(k'-k+i\epsilon)} G_{+}(k')\left(\frac{1}{2\sinh^2\left(\frac{k'}{2}\right)}+F(k')\right)e^{-i\mu k'}\\-\int_{-\infty}^{\infty}\frac{dk'}{2\pi i(k'-k+i\epsilon)}G_{+}(k')X_{-}(k')e^{-2i\mu k'}~.
\end{multline}
In the regime of very large, but finite $\mu$, the last term can be argued to be small. Treating this term
perturbatively leads to an approximation scheme where we have an accurate description only at the right endpoint of the
eigenvalue density. However, since we add only even sources to the action we can find a valid description for $\rho(x)$
in the interval $[0,\mu]$, assume that $\rho(x)$ is an even function, and reflect it around $x=0$. In this way, one
finds \cite{Passerini:2011fe}
\begin{equation}
\rho(k)=-G_{-}(k)e^{i k\omega}\int^{\infty}_{-\infty}\frac{dk'}{2\pi i(k'-k+i\epsilon)} G_{+}(k')\left(\frac{1}{2\sinh^2\left(\frac{k'}{2}\right)}+F(k')\right)e^{-i\mu k'}~.
\end{equation}
The integral has the effect of subtracting the poles in the lower half-plane, therefore this expression is of the form
(\ref{FourierExample}). The integral that appears in this equation can be evaluated further by closing the contour
around the lower half-plane
\begin{equation}
\rho(k)=\frac{1}{\cosh(k)}+\frac{2\sinh^2(\frac{k}{2})}{\cosh(k)}F(k)+G_{-}(k)e^{i\mu k}\sum_{l=0}^\infty\frac{r_l e^{-\mu \nu_l}}{k+i\nu_l}\left(1-F(-i\nu_l)\right)~.
\label{strongdensity}
\end{equation}
At the first iteration of this scheme
\begin{multline}
F(k)=\frac{1}{2\sqrt{\pi}}\sum_{n=1}n g_{2n}\mu^{2n}\pi^n\sum_{p=0}^{n-1}b_{n-p-1}\frac{\Gamma(p+\frac{1}{2})}{\Gamma(p+2)}\,_1F_2(n+\frac{1}{2};\frac{1}{2},p+2;-\frac{1}{4}k^2\mu^2)~.
\end{multline}

\subsection{Computing the moments}

The condition that arises from the normalization of the eigenvalue distribution was found in \cite{Passerini:2011fe} by
exploiting the fact that the eigenvalue density is an even function
\begin{equation}
1=2\int_{0}^{\mu}dx\, \rho(x)=\int_{-\infty}^{\infty}\frac{dk}{2\pi}\frac{\rho(k)}{k-i\epsilon}
~.\end{equation}
By closing the contour in the lower half-plane, \cite{Passerini:2011fe} finds
\begin{equation}
\sum_{l=0}^{\infty}\frac{r_{l}e^{-\mu\nu_l}F(-i\nu_{l})}{\nu_l+\nu_0}=\frac{r_{0}}{2\nu_0}e^{-\mu\nu_0}.
\label{StrongNormalisation}
\end{equation}

Moreover, we have the eigenvalue density in Fourier space we can find an expression for the moments as follows
\begin{multline}
\label{m2n}
m_{2n}=2\int_{0}^{\mu}dx\,x^{2n}\int_{-\infty}^{\infty}\frac{dk}{2\pi} \, e^{-ikx}\rho(k)=\frac{i^{2n}}{\pi}\int dk\, \rho(k)\int_{0}^{\mu}\frac{d^{2n}}{dk^{2n}} e^{-ikx}\\=\frac{i^{2n+1}}{\pi}\int_{-\infty}^{\infty}dk\, \rho(k)\frac{d^{2n}}{dk^{2n}}\frac{e^{-ik\mu}-1}{k}\simeq \frac{(-1)^n(2n)!}{i\pi}\int_{-\infty}^{\infty}dk\, \frac{\rho(k)}{k^{2n+1}}~.
\end{multline}
In the last step we have eliminated all terms that oscillate rapidly in the $\mu\gg 1$ regime. From the form of
$\rho(k)$ it is clear that we can close the contour in the positive half-plane and pick up all the poles, which are the
ones originating from $G_-(k)$ and the higher order poles at zero. Let us consider the manipulations of each of the
terms appearing in \eqref{m2n} individually.

First, the term
\begin{equation}
\frac{(-1)^n(2n)!}{i\pi}\int_{-\infty}^{\infty}dk\, \frac{2\sinh^2(\frac{k}{2})}{k^{2n+1}\cosh(k)}F(k)
\end{equation}
is simply given by evaluating the residue at zero
\begin{equation}
(-1)^n\lim_{k\rightarrow 0}\frac{d^{2n}}{dk^{2n}}\frac{2\sinh^2(\frac{k}{2})}{\cosh(k)}F(k)~.
\end{equation}
The last term is only slightly more complicated
\begin{multline}
\frac{(-1)^n(2n)!}{i\pi}\int_{-\infty}^{\infty}\frac{dk}{k^{2n+1}}G_{-}(k)e^{i\mu k}\sum_{l=0}^{\infty}\frac{r_le^{-\mu\nu_l}}{k+i\nu_l}(1-F(-i\nu_l))\\=(-1)^n\lim_{k\rightarrow 0}\frac{d^{2n}}{dk^{2n}}G_-(k)e^{i\mu k}\sum_{l=0}^{\infty}\frac{r_le^{-\mu\nu_l}}{k+i\nu_l}(1-F(-i\nu_l))\\+2(-1)^n(2n)!\sum_{l,m=0}^{\infty}\frac{r_l r_m e^{-\mu(\nu_l+\nu_m)}}{(i\nu_m)^{2n+1}(i\nu_l+i\nu_m)}(1-F(-i\nu_l))~.
\end{multline}
By noting that $\lim_{k\rightarrow 0}G_{-}(k)=0$, the first line of the r.h.s.\ can be simplified to
\begin{multline}
(-1)^n\lim_{k\rightarrow 0}\frac{d^{2n}}{dk^{2n}}G_-(k)e^{i\mu k}\sum_{l=0}^{\infty}\frac{r_le^{-\mu\nu_l}}{k+i\nu_l}(1-F(-i\nu_l))\\=(-1)^n\left[\frac{d^{2n}}{dk^{2n}}(-i)G_-(k)\right]_{k=0}\sum_{l=0}^{\infty}\frac{r_le^{-\mu\nu_l}}{\nu_l}(1-F(-i\nu_l))~.
\end{multline}
For the last line, note that it is dominated by the first term in the sum over $m$
\begin{multline}
\sum_{l,m=0}^{\infty}\frac{r_l r_m e^{-\mu(\nu_l+\nu_m)}}{(i\nu_m)^{2n+1}(i\nu_l+i\nu_m)}(1-F(-i\nu_l))\simeq
\sum_{l=0}^{\infty}\frac{r_l r_0 e^{-\mu(\nu_l+\nu_0)}}{(i\nu_0)^{2n+1}(i\nu_l+i\nu_0)}(1-F(-i\nu_l))=0,
\end{multline}
and is therefore killed due to the normalization condition (\ref{StrongNormalisation}). 

Putting everything together we find the following expression for the moments
\begin{equation}
m_{2n}=h_n+(-1)^n\lim_{k\rightarrow 0}\frac{d^{2n}}{dk^{2n}}\frac{2\sinh^2(\frac{k}{2})}{\cosh(k)}F(k)+(-1)^n\left[\frac{d^{2n}}{dk^{2n}}(-i)G_-(k)\right]_{k=0}\sum_{l=0}^{\infty}\frac{r_le^{-\mu\nu_l}}{\nu_l}(1-F(-i\nu_l)),
\label{moments1}
\end{equation}
where $h_n$ is an unimportant set of constants. We want to calculate correlators which are derivatives of the moments
with respect to coupling constants, so the explicit value of the constants $h_{n}$ is not needed.

Next let us focus on the infinite sum in the rightmost term of the last equation, \eqref{moments1}. We rewrite this term
by means of the normalization condition \eqref{StrongNormalisation} and by removing terms that are exponentially
suppressed in the regime $\mu \gg 1$,
\begin{multline}
\sum_{l=0}^{\infty} \frac{r_{l}e^{-\mu \nu_l}}{\nu_l}\left(1-F(-\nu_l)\right)=\frac{r_0e^{-\mu \nu_l}}{\nu_0}-\sum_{l=0}^{\infty}\frac{r_l e^{-\mu \nu_l}}{\nu_l}F(-i\nu_l)\\=\sum_{l=0}^{\infty} \frac{2r_l e^{-\mu \nu_l}}{\nu_l+\nu_0}F(-i\nu_l)-\frac{r_le^{-\mu\nu_l}}{\nu_l}F(-i\nu_l)=\sum_{l=0}^{\infty}\frac{(\nu_l-\nu_0)r_le^{-\mu\nu_l}}{\nu_l(\nu_l+\nu_0)}F(-i\nu_l)~.
\end{multline}
Continuing to ignore the exponentially suppressed terms the hypergeometric function contained within $F(-i\nu_{l})$ has
the following asymptotic expansion
\begin{equation} 
\,_1F_2\left(n+\frac{1}{2};\frac{1}{2},n+2;\frac{\nu_l^2\mu^2}{4}\right)\sim\frac{\Gamma(n+2)}{\Gamma(n+\frac{1}{2})}e^{\mu\nu_l}\sum_{k=0}^{\infty}\sqrt{2}c_k(\mu\nu_l)^{-k-\frac{3}{2}}~,
\end{equation}
where the coefficients $c_{k}$ obey the recursive equation
\begin{equation}
c_{k}=-\frac{1}{8k}(1+2k)(7-6k+8n)c_{k-1}-\frac{1}{16k}(1+2k)(2k-1)(2k-4n-5)c_{k-2}
\end{equation}
with $c_0=1$ and $c_1=-3n-\frac{3}{8}$. Now expanding each term of $F(-i\nu_l)$ 
up to the order of interest we obtain
\begin{multline}
\sum_{l=0}^{\infty}\frac{(\nu_l-\nu_0)r_le^{-\mu\nu_l}}{\nu_l(\nu_l+\nu_0)}F(-i\nu_l)=\frac{1}{\sqrt{2\pi}}\sum_{n=1}ng_{2n}\sum_{p=0}^{n-1}b_{n-p-1}\sum_{k=0}^{2n-2}c_k \alpha_k \pi^{n-k-\frac{3}{2}} \mu^{2n-k-\frac{3}{2}}~,
\end{multline}
where $\alpha_k$ is a set of numerical constants defined as
\begin{equation}
\alpha_k\equiv\sum_{l=0}^{\infty}\frac{lr_{l}}{l+1}(l+\frac{1}{2})^{-k-\frac{5}{2}}~.
\end{equation}
This series converges and can therefore be determined up to arbitrary numerical precision by computing partial sums. 

Including this result into the expression \eqref{moments1} we find the following expression for the moments 
\begin{multline}
m_{2n}=h_n+(-1)^n\lim_{k\rightarrow 0}\frac{d^{2n}}{dk^{2n}}\frac{2\sinh^2(\frac{k}{2})}{\cosh(k)}F(k)\\+(-1)^n\frac{1}{\sqrt{2\pi}}\left[\frac{d^{2n}}{dk^{2n}}(-i)G_-(k)\right]_{k=0}\sum_{n=1}ng_{2n}\pi^{n}\sum_{p=0}^{n-1}b_{n-p-1}\sum_{k=0}^{2n-2}c_{k}\alpha_{k}\pi^{n-k-\frac{3}{2}}\mu^{2n-k-\frac{3}{2}}~.
\end{multline}
This formula gives the moments at the first iteration of the above scheme, where the last term on the r.h.s. of equation
\eqref{rhoequation} has been completely dropped.

\subsection{Computing correlators}
Unlike the computation of correlators in the weak coupling regime, in this subsection we will bypass the evaluation of
the planar free energy by means of the following relation
\begin{equation}
\frac{\partial}{\partial g_{2n}}F=\pi^{n}\int_{-\mu}^{\mu}dz\, z^{2n}\rho(z)=\pi^{n}m_{2n}~.
\end{equation}
Since the correlators are, by default, given by higher derivatives of the free energy we can obtain them by taking
derivatives of the moments \cite{Rodriguez-Gomez:2016ijh}
\begin{multline}
\left<\Tr[\phi^2]^{n_2}\Tr[\phi^3]^{n_3}...\Tr[\phi^m]^{n_m},\Tr[\bar{\phi}^2]^{\bar{n}_2}...\Tr[\bar{\phi}^m]^{\bar{n}_m}\right>_{S^4}\\=-N^{2}\prod_k \left(\frac{\partial}{\partial g_k}\right)^{n_k+\bar{n}_k}F(\lambda,g_{n})=-N^{2}\prod_{k\neq i}\pi^{i} \left(\frac{\partial}{\partial g_k}\right)^{n_k+\bar{n}_k}\left(\frac{\partial}{\partial g_i}\right)^{n_k+\bar{n}_k-1}m_{2i}~.
\end{multline}
This relation also provides a non-trivial check for the moments as they naturally have to satisfy the following
commutativity property $\pi^{m}\frac{\partial}{\partial g_{2n}}m_{2m}=\pi^{n}\frac{\partial}{\partial g_{2m}}m_{2n}$.

\begin{figure}
	\centering
		\includegraphics[width=0.60\textwidth]{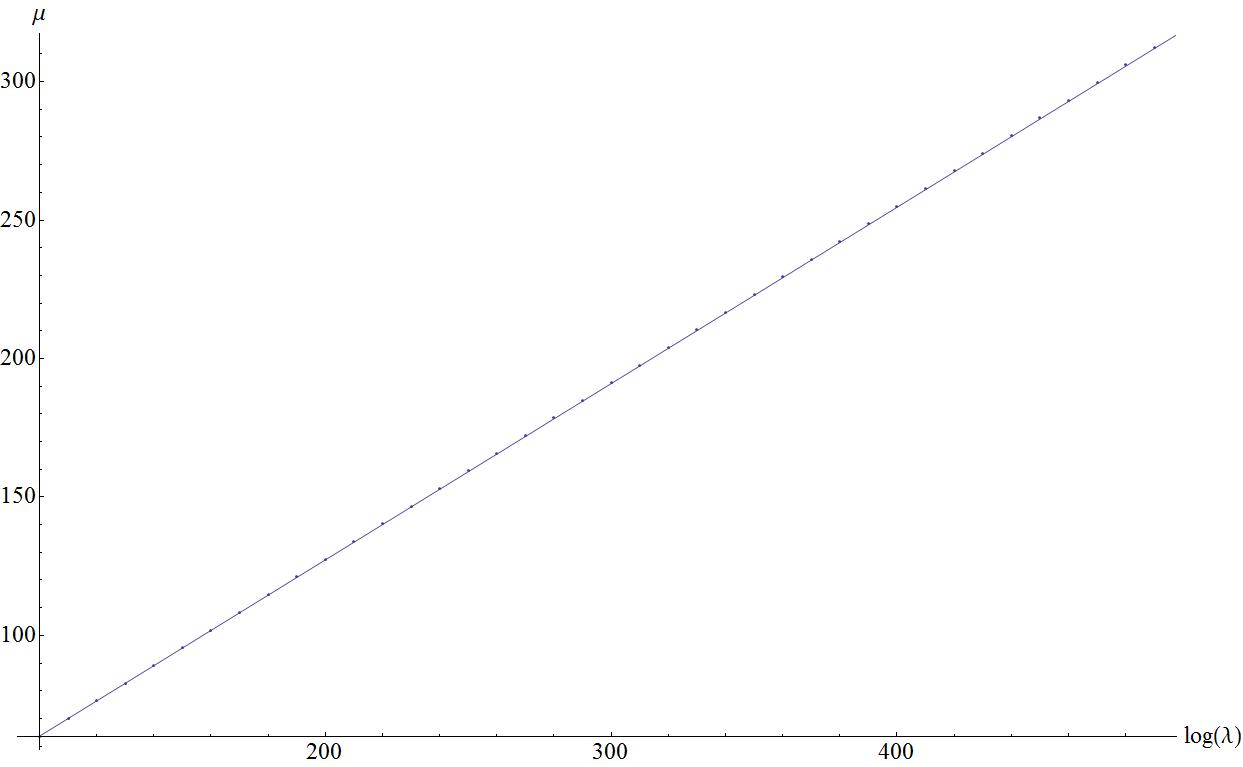}
	\caption{The dotted line represents the numerical solution of $\mu$ as a function of $\log(\lambda)$ after all other source terms have been set to zero, for comparison the solid line represents a plot of $\mu=\frac{2}{\pi}\log(\lambda)$}
	\label{musolvedplot}
\end{figure}

As in the weak coupling computation, one of the most difficult steps is the inversion of the normalization condition.
In its present form it is an implicit function relating the size of the cut $\mu$ to the coupling constants $g_{2n}$. We
should use this to eliminate $\mu$ from the correlators in favor of the couplings. The normalization condition is given
by equation (\ref{StrongNormalisation}). The plot of the numerical solution of the normalization condition as a function
of $\log(\lambda)$ (see Fig.\ (\ref{musolvedplot})) suggests that $\mu\sim\frac{2}{\pi}\log(\lambda)$ in the strong
coupling regime where $\lambda\gg 1$. This scaling was already conjectured in \cite{Passerini:2011fe} by means of an
interpolation between the weak coupling and infinite 't Hooft coupling regime.

Using this proposed behavior for $\mu$ as a function of $\lambda$ and the orthogonalization procedure of
\cite{Gerchkovitz:2016gxx} we find the leading order behavior for the correlators quoted in section \ref{strong}.

\end{appendix}

\bibliography{ttstarpaper}
\bibliographystyle{utphys}

\end{document}